\documentstyle[prb,aps,twocolumn,psfig]{revtex}
\begin{document}
\draft

\title{Superconducting states and depinning transitions of Josephson ladders}
\author{Mauricio Barahona\thanks{Present address:
Ginzton Laboratory, Stanford University, Stanford, CA 94305.}}
\address{Physics Department, Massachusetts Institute of Technology,
Cambridge, MA 02139}
\author{Steven H. Strogatz}
\address{Kimball Hall, Department of Theoretical and Applied Mechanics and
Center for Applied Mathematics, Cornell University, Ithaca, NY 14853}
\author{Terry P. Orlando}
\address{Department of Electrical Engineering and Computer Science,
         Massachusetts Institute of Technology,
         Cambridge, MA 02139}
\date{To appear in Phys.\ Rev.\ B, Jan 1 1998}
\maketitle

\begin{abstract}
We present analytical and numerical studies of pinned superconducting
states of open-ended Josephson ladder arrays, neglecting inductances but
taking edge effects into account.  Treating the edge effects
perturbatively, we find analytical approximations for three of these
superconducting states -- the no-vortex, fully-frustrated and single-vortex
states -- as functions of the dc bias current $I$ and the frustration $f$.
Bifurcation theory is used to derive formulas
for the depinning currents and critical frustrations at which the
superconducting states disappear or lose dynamical stability as $I$ and $f$
are varied.  These results are combined to yield a zero-temperature
stability diagram of the system with respect to $I$ and $f$.  To highlight
the effects of the edges, we compare this dynamical stability diagram to
the thermodynamic phase diagram for the
infinite system where edges have been neglected. We briefly
indicate how to extend our methods to include self-inductances.
\end{abstract}


\section{Introduction}

Arrays of Josephson junctions are of interest in several branches of
physics~\cite{scbooks}.  They have many technological
applications, including high-frequency emitters and detectors, parametric
amplifiers, local oscillators, and voltage
standards~\cite{scbooks,technology}.  They also shed
light on the structural~\cite{kleiner} and pinning~\cite{mpa1}
properties of the high-T$_{\rm c}$ superconducting cuprates. At the same
time, they provide model systems for the study of problems in
both spatiotemporal nonlinear
dynamics~\cite{series,shiseries,shilong,summary} and
nonequilibrium statistical physics~\cite{summary,kardar0}.  For instance,
the depinning transitions and nonlinear wave propagation seen in Josephson
arrays are analogous to those found in incommensurate systems, earthquake
models, Type II superconductors, and charge-density waves.

From the standpoint of dynamical systems theory~\cite{steve}, Josephson
arrays can be viewed as large collections of coupled nonlinear oscillators.
Unfortunately, because of their nonlinearity and large number of degrees of
freedom, these arrays are inherently difficult to analyze
mathematically.  A further complication is that there is an intrinsic
physical coupling among junctions, due to fluxoid quantization, which is more
awkward to handle than the nearest-neighbor interaction usually assumed in
idealized models of coupled oscillators.  And when the effects of 
self-fields and inductances are
included, there is even less hope of making analytical progress.

Despite these obstacles, some encouraging advances have occurred recently
in the mathematical analysis of Josephson arrays, especially for
one-dimensional (1-D) systems where the junctions are connected in
series~\cite{series,shiseries} or in parallel~\cite{shilong}.  The logical
next step is to tackle two-dimensional (2-D) arrays.  Much of the previous
theoretical work on 2-D arrays has focused on numerical simulation of the
current-voltage characteristics~\cite{experIV}, in an effort to link the
rich spatiotemporal dynamics of 2-D arrays to the averaged quantities that
are most readily measured experimentally.  On the mathematical side, there
are recent indications that 2-D arrays, like their 1-D counterparts, are
also going to be tractable in some regimes~\cite{analysis2d}.

An ideal example to explore the crossover between 1-D and
2-D behavior is the Josephson ladder array (Figure~\ref{fig:ladder}).
Following Kardar~\cite{kardar1,kardar2}, several authors have
studied various statistical properties of the frustrated ladder, including
its ground state, the complicated landscape of solutions at zero
temperature, the low-lying excitations, and the linear response
regime~\cite{benedict,falo2,denniston}. However, all of these authors
restricted attention to ladders in the absence of a driving current.  Only
recently has the fully dynamical problem been addressed, through numerical
simulations of the depinning transition~\cite{stroudscreen} and vortex
propagation~\cite{strouddyn}.

In this paper we use the tools of nonlinear dynamics to analyze 
the superconducting states of ladder arrays. (Other dynamical
regimes will be discussed elsewhere~\cite{mythesis,usladder2}.)
Mathematically, the superconducting states correspond to fixed points of
the governing circuit equations.  We study the bifurcations of these fixed
points with respect to variations in the applied dc bias current $I$ and
the frustration $f$ introduced by an external magnetic field.  As $I$ is
increased from $0$ at fixed $f$, we find that the stable superconducting
states are destroyed in saddle-node bifurcations at certain critical values
of the current.  Then the system depins from its original static
configuration and evolves toward some other state.  The new state might be
another fixed point, or it might be a running solution, in which case a
nonzero dc voltage appears across the array.  Global depinning of the array
occurs when the last stable fixed point is destroyed.  One of the main
results of our analysis is a set of approximate analytical formulas for the
critical currents at which the fixed points are destroyed, as a function of
$f$, for the three most important types of superconducting states: the
no-vortex, fully-frustrated and single-vortex solutions.

Another important finding is that symmetry plays a crucial role in the
dynamics of the ladder.  As we will show below, much of the behavior of the
ladder can be understood by focusing on states that are ``up-down symmetric''
-- in other words, states where the phases of the horizontal junctions on
the top and bottom of any given plaquette are equal in magnitude but
opposite in sign at all times.  All of the stable superconducting states
possess this symmetry.  Even when the parameters are chosen so that
depinning occurs, the subsequent transients and long-term running solutions
typically remain up-down symmetric.  But there is at least one
exception: when the single-vortex state is destabilized by lowering the
frustration $f$ below some critical value $f_{\rm min}$, the system depins
via a symmetry-breaking bifurcation.  During the transient behavior, the
up-down symmetry is lost temporarily, but is then recovered as the system
expels flux from the array and evolves toward the no-vortex state.

These symmetry considerations establish an unexpected link between the
study of Josephson arrays and some recent developments in nonlinear
dynamics.  In mathematical terms, the up-down symmetric states of a ladder
array form an invariant manifold of the full state space.  A
symmetry-breaking bifurcation occurs when this manifold loses stability in
a transverse direction.  The same issue -- the transverse stability of an
invariant manifold -- arises in the study of riddled basins, synchronized
chaos, on-off intermittency, and blowout bifurcations~\cite{transverse}.
These connections suggest a promising line of future research on Josephson
arrays, particularly with regard to their chaotic states.

We will also show that the dynamics of ladder arrays are very strongly
influenced by edge effects.  One might have supposed these effects to be
negligible, especially in long ladders, since their influence on the
superconducting solutions dies off exponentially fast away from the
boundaries.  Yet although the edges do indeed have a small effect on the
{\it form} of the superconducting solutions, they have a large effect on
the {\it stability} of those solutions.  Much of this paper is devoted to
investigating the effects of the edges, first on the superconducting states
themselves, then on their stability, and finally on the entire phase
diagram.

This paper is organized as follows.  Section~\ref{sec:system}
reviews the model equations for the ladder and discusses their
symmetry properties.  In Section~\ref{sec:analytical}
we obtain analytical approximations for three numerically observed
superconducting solutions: the no-vortex (nv), single-vortex (sv),
and fully-frustrated (ff) solutions.  In all cases, edge effects are
taken into account via perturbation theory.
Next, in Section~\ref{sec:depinning} we describe the dynamical simulations
which reveal the depinning properties of the nv, sv, and ff configurations
and relate them to the global depinning of the ladder.
In Section~\ref{sec:bifurc} we establish the rigorous connection of the
dynamical depinning with the stability of these three fixed points.
For all of them, we characterize the bifurcations and study their
stability diagrams, both in the presence and in the absence of edges.
When possible, analytical approximations to the critical
currents are obtained.  We show that the depinning transitions correspond
to saddle-node bifurcations that are edge-dominated
for almost all values of the frustration.
Moreover, we find that some of the superconducting states can be
destabilized via a subcritical pitchfork bifurcation as the frustration is
reduced; in physical terms, this is a symmetry-breaking bifurcation in
which flux is expelled {\it transversally} from the ladder.  In the final 
section we summarize our conclusions in two phase diagrams
(for ladders with and without edges), and we
relate our results to those found by previous authors. We also add two
more technical appendices:
Appendix A compares the single-vortex configuration in the ladder with the
corresponding kink-like solution in 1-D parallel arrays; Appendix B
briefly indicates how to extend our approach to include self-inductance
effects.

\section{The system}
\label{sec:system}
\begin{figure}[t]
\begin{center}
\hspace*{.1in}
\psfig{file=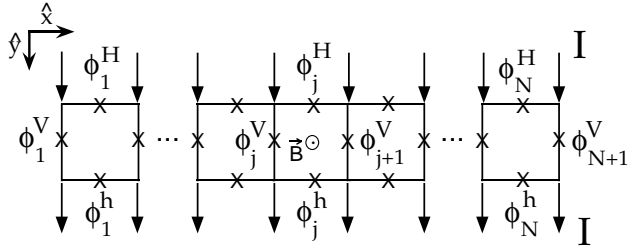,width=3.5in}
\end{center}
\caption{Schematic diagram of the Josephson ladder array with dc current
$I$ injected in the perpendicular (${\bf \hat y}$) direction.
The external magnetic field  ${\bf B}$ is applied transversely to the
plane of the device, in the ${\bf -\hat z}$ direction.}
\label{fig:ladder}
\end{figure}

We study an open-ended Josephson ladder with $N$ square plaquettes, i.e., an
array formed by two rows of $N+1$ weakly coupled superconducting islands
(Figure~\ref{fig:ladder}). The array is
driven by a perpendicular uniform dc current $I$, and a magnetic field is
applied transverse to the plane of the device.  Each weak link between
islands constitutes a junction.  Its state is described by the
gauge-invariant phase
difference $\phi_{j}$, arising from the macroscopic character of the
quantum wavefunction of the superconductors.

Assuming zero temperature, negligible charging (quantum) effects, and
identical junctions, the dynamics of each junction is given, in the
three-channel RCSJ
model~\cite{scbooks}, by the nonlinear differential equation
\begin{equation}
\label{eq:pendulum}
I_{j}={\ddot \phi_{j}} + \beta_{c}^{-1/2}
{\dot \phi_{j}} + \sin \phi_{j} \equiv {\cal N}(\phi_{j}).
\end{equation}
Here $\beta_{c}$ is the McCumber parameter~\cite{scbooks};
$I_{j}$ is given in units of the critical current of each (identical) junction;
derivatives are with respect to time normalized in units of $\omega_{J}^{-1}$,
the inverse of the plasma frequency;
and ${\cal N}$ is shorthand for the nonlinear differential operator defined
by~(\ref{eq:pendulum}).
Thus, each junction is formally equivalent to a damped driven
pendulum~\cite{steve}.

The junctions are {\it intrinsically} coupled, even when
inductances are neglected, through two physical restrictions: the
quantization of the magnetic flux through each plaquette,
and Kirchhoff's current conservation law at each node.
When all inductances are zero, i.e., self-fields are neglected, the flux
quantization condition for the array in Figure~\ref{fig:ladder} becomes
\begin{equation}
\label{eq:fluxoid}
\phi_{j}^{V}+\phi_{j}^{h}-\phi_{j+1}^{V}-\phi_{j}^{H}=2 \pi (n_{j}-f),
\hspace{.1in}  j=1,\ldots,N
\end{equation}
where $f$ is the {\it external} magnetic flux in units of the flux quantum
$\Phi_{0}$. The set of integers $\{n_{j}\}$ indicate the presence
($n_{j}=\pm 1$) or absence ($n_{j}=0$) of topological vortices in each
plaquette when the phases are restricted to the interval $[-\pi,\pi)$.

In addition, Kirchhoff's current law yields
\begin{eqnarray}
\label{eq:interiorequation1}
&I_{j-1}^{H} + I = I_{j}^{H} + I_{j}^{V},\;\;& j=2,\ldots,N\\
&I_{j-1}^{h} + I_{j}^{V} = I + I_{j}^{h},\;\;& j=2,\ldots,N
\label{eq:interiorequation2}
\end{eqnarray}
at the interior nodes of the ladder.  At the left edge,
\begin{eqnarray}
\label{eq:edgeequation1}
&I = I_{1}^{H} + I_{1}^{V}, \\
&I_{1}^{V} = I + I_{1}^{h},
\label{eq:edgeequation2}
\end{eqnarray}
while at the right edge,
\begin{eqnarray}
\label{eq:edgeequation3}
&I_{N}^{H}+ I =  I_{N+1}^{V}, \\
&I_{N+1}^{V} + I_{N}^{h} = I .
\label{eq:edgeequation4}
\end{eqnarray}
In summary, equations~(\ref{eq:pendulum})--(\ref{eq:edgeequation4}) define
our model for the dynamics of the ladder array.  A useful mechanical analog
for the system is a frustrated lattice of coupled, damped, nonlinear
pendula driven by a constant torque applied at the edges.

An important restriction on the currents immediately follows from the
presence of the edges.  Equations~(\ref{eq:edgeequation1})
and~(\ref{eq:edgeequation2}) imply that
$I_{1}^{H} = -I_{1}^{h}$.  Moving successively from the left edge to the
interior of the ladder, Kirchhoff's current law yields
\begin{equation}
\label{eq:returnedges}
I_{j}^{H} = -I_{j}^{h},
\hspace{.2in}  \forall \; j.
\end{equation}

This condition~(\ref{eq:returnedges}) is automatically satisfied by any
phase configuration whose evolution obeys the {\it up-down symmetry}
\begin{equation}
\label{eq:phiupdown}
\phi^{H}_{j}(t)=-\phi^{h}_{j}(t), \hspace{.3in} \forall \; j, \,
\forall \; t
\end{equation}
as can be seen from~(\ref{eq:pendulum}).
Moreover, if the initial conditions satisfy
$\phi^{H}_{j}(0)=-\phi^{h}_{j}(0)$ along with similar equalities on the first
time-derivatives, the governing equations imply that those
equalities will hold for {\it all} time. In geometrical terms,
the set of all up-down symmetric states~(\ref{eq:phiupdown}) forms an
invariant submanifold of the full phase space.

On the other hand, it is certainly possible to choose initial conditions
that do not have this up-down symmetry.  But our simulations indicate that
for a wide range of parameters and initial conditions,  arbitrary phase
configurations rapidly evolve toward up-down symmetric states.  In other
words, the invariant manifold is typically attracting in the transverse
directions --- initial states that are off the manifold are soon drawn onto
it.  There are also exceptions to this rule: as we will see in
Section~\ref{sec:bifurc}, the single-vortex and fully-frustrated states
can lose transverse stability as the frustration $f$ decreases.
Nevertheless, a great deal of
insight can be obtained by restricting attention to the submanifold of
up-down symmetric states.  Thus, for much of this paper we will assume
that~(\ref{eq:phiupdown}) holds, and we will replace
$\phi^{h}_{j}$ with $-\phi^{H}_{j}$ throughout the governing equations.

There is a simple condition for the transverse stability of an
up-down symmetric fixed point.
It can be shown~\cite{mythesis} that such a fixed point is
linearly stable to perturbations that are {\it strictly normal} to
the manifold if and only if
\begin{equation}
\label{eq:horizontalcond}
|\phi^{H}_{j}| < \pi /2,
\hspace{.2in}  \forall \; j.
\end{equation}
We have checked numerically that in the instances when the up-down
symmetry is broken (and, thus, the system escapes the symmetric manifold),
the subsequent evolution does in fact take place purely along the normal
direction.
Therefore, any up-down symmetric fixed point that is stable must satisfy
this inequality~(\ref{eq:horizontalcond}).  Hence~(\ref{eq:horizontalcond})
constitutes a {\it necessary} condition for stability.  (It is not
sufficient, however,
because it only governs the transverse direction; it says nothing about the
stability with respect to perturbations that preserve the symmetry.
Further conditions would be needed to ensure stability in directions {\it
along} the invariant manifold as will be shown below.)

In summary, the governing equations can be written compactly as
${\bf f(x) = 0}$, with
${\bf x}=(\phi^{V}_{1},\ldots,\phi^{V}_{N+1},\phi^{H}_{1},\ldots,
\phi^{H}_{N},\phi^{h}_{1},\ldots,\phi^{h}_{N})$, and with the components 
of ${\bf f(x)}$ defined by
\begin{eqnarray}
\label{eq:eq1}
{\rm f}_{j}({\bf x})&=& I+{\cal N}(\phi^{H}_{j-1})-{\cal N}(\phi^{H}_{j})-
{\cal N}(\phi^{V}_{j}), \nonumber\\
&& \hspace*{1in}j=1, \ldots, N+1\\
 {\rm f}_{N+1+j}({\bf x})&=&{\cal N}(\phi^{H}_{j})+{\cal N}(\phi^{h}_{j}), 
\nonumber\\
&& \hspace*{1in}j=1, \ldots, N\\ 
\label{eq:eq2}
 {\rm f}_{2N+1+j}({\bf x})&=&\phi_{j}^{V}+\phi_{j}^{h}-\phi_{j+1}^{V}
-\phi_{j}^{H}-2 \pi(n_{j}- f), \nonumber\\
&& \hspace*{1in}j=1,\ldots,N.
\label{eq:eq3}
\end{eqnarray}
The dynamical evolution of the state $\{{\bf x}(t),{\dot {\bf x}}(t)\}$
of the system is obtained by
numerically solving this system of coupled
differential and algebraic equations.  The dynamics also depend implicitly
on the parameters $I, \beta_{c}$, $f$, and $N$.

In experiments, the most convenient way to probe the dynamics of the array
is to measure its dc current-voltage ($IV$) characteristics.
 From the Josephson relations~\cite{scbooks}, the
time-dependent voltage across each junction is directly proportional
to ${\dot{\phi}_{j}}$, the time-derivative of its phase.  Hence, the total dc
voltage $V$ across the array in the vertical direction is proportional to
the spatial {\it and} temporal average of all the phase derivatives.
Although because of this averaging a great deal of dynamical
information is lost about the spatiotemporal state of the system,
the $IV$ curve still provides a useful (if somewhat coarse)
indicator of changes in the underlying dynamics as the drive current $I$ is
varied.

In the case of ladder arrays, the $IV$ curves display three regions
associated with distinct dynamical behaviors~\cite{strouddyn,mythesis}.
At low $I$, the system is {\it superconducting}
($V=0$) with pinned or slightly oscillating junctions (in the mechanical
analog, the pendula are at rest or librating slightly).
At a depinning current $I_{\rm dep}$, the array jumps to the
{\it flux-flow} region, in which a finite voltage is produced by
vortices of magnetic flux moving across the array.
At still higher currents, the
dynamics is characterized by {\it whirling modes}~\cite{shilong} (in the
mechanical
analog,  all the pendula rotate over the top at a nearly uniform angular
velocity proportional to the applied torque). In this state, there is a
linear {\it ohmic} dependence of $V$ on $I$.
In the remainder of the article we focus on the superconducting
states and the critical current $I_{\rm dep}$ at which depinning occurs.

\section{Observed superconducting solutions and analytical approximations}
\label{sec:analytical}

We have performed dynamical simulations of the array in which the
current is ramped up from zero with different initial conditions.
The superconducting  solutions observed in those simulations are always
static states, i.e., fixed points of the system.  (In principle,
time-dependent solutions with high cancelling symmetry could also have zero
total dc voltage, but we never see such states in our simulations.)   More
specifically, for any given $\beta_{c}$ and $N$, only three types of
configurations appear in the numerics:
{\it no-vortex} solutions (Figure~\ref{fig:scnovortex})
for $f$ smaller than $\sim 0.3$, and solutions of the
{\it single-vortex} type (Figure~\ref{fig:onevortex}) and
{\it fully-frustrated} type (Figure~\ref{fig:schalfedge})
for $f \rightarrow 1/2$.
Far from the edges, the no-vortex (nv) state  is
characterized by identical phases for all junctions.
The same applies to the single-vortex (sv) configuration far
from both the edges and the center of the vortex. On the other hand, the
fully-frustrated (ff) state has a spatial oscillatory pattern with a
wavelength equal to two plaquettes. All of these states are modified by
noticeable edge effects.
Although there are many other static
solutions of the system, our numerical simulations indicate that
the no-vortex, single-vortex, and fully-frustrated states are the only ones
needed to explain the depinning behavior of the array.

When dealing with {\it fixed points} with up-down symmetry~(\ref{eq:phiupdown})
the defining equations~(\ref{eq:eq1})--(\ref{eq:eq3}) become
\begin{eqnarray}
\label{eq:scequation1}
&I+\sin \phi^{H}_{j-1}=\sin \phi^{H}_{j}+\sin \phi^{V}_{j},\; \;
&j=1,\ldots,N+1 \\
&\phi^{V}_{j}-\phi^{V}_{j+1}-2 \phi^{H}_{j}= 2 \pi (n_{j}-f), \; \;
&j=1,\ldots,N
\label{eq:scequation2}
\end{eqnarray}
where we have defined artificial phases $\phi^{H}_{0}=\phi^{H}_{N+1}=0$.
Note that the McCumber parameter $\beta_{c}$ does not appear in the
equations for the fixed points, and hence does not affect their existence.
This is consistent with the numerically observed independence of the
depinning behavior on $\beta_{c}$.

In the rest of this section we obtain analytical approximations for the
nv, sv, and ff configurations mentioned above. We follow a common
scheme for all of them. First, we obtain a no-edge approximation
(denoted with a dagger $\dagger$) for the infinite ladder. Second, we
introduce the effect of the edges perturbatively to obtain an
edge-corrected approximation (identified by a double dagger $\ddagger$).
The calculated configurations have been exhaustively compared with
the results of numerical simulations with excellent agreement.

\subsection{No-vortex solution}

\begin{figure}[t]
\psfig{file=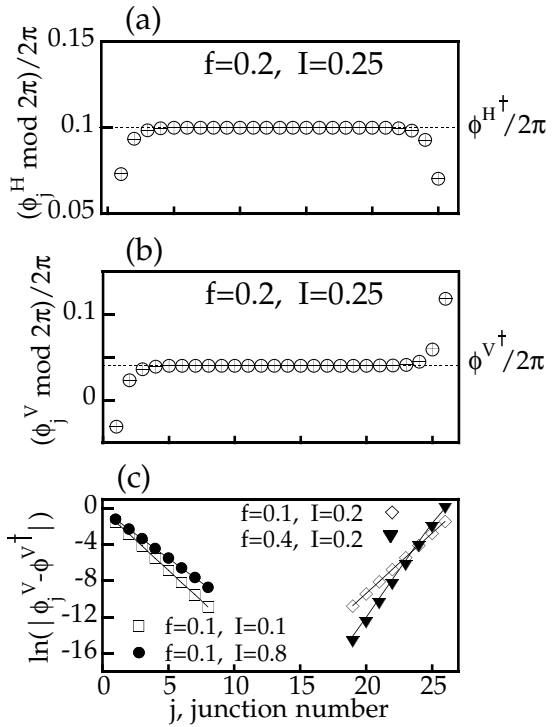,width=3.5in}
\caption{ No-vortex superconducting solution. (a) Phases of the
horizontal junctions for a $25 \times 1$ array with $I=0.25$ and $f=0.2$:
numerically observed solution (+), and approximate solution
${\phi^{H}_{j}}^{\ddagger}$ ($\circ$). (b) Same as in (a) for the
vertical junctions. The predicted phases of the infinite-ladder solution
$\{{\phi^{V}_{j}}^{\dagger},{\phi^{H}_{j}}^{\dagger}\}$ (short-dashed line)
are seen to be valid far from the edges in both (a) and (b).
(c) Exponential decay of the correction from the edges with varying
magnetic field $f$ and current $I$. The symbols correspond to the
numerically calculated vertical phases. The solid lines come from our
approximate solution ${\phi^{V}_{j}}^{\ddagger}$
and have slopes $\pm \ln r$ given by~\protect{(\ref{eq:scnvr})}.}
\label{fig:scnovortex}
\end{figure}

Figure~\ref{fig:scnovortex} (a)--(b) shows a plot of the no-vortex solution, as
computed numerically, along with the analytical approximation described
below. This state is characterized by the absence of
topological vortices ($n_{j}=n_{j+1}=0, \; \forall j$)
and, far from the edges, by the constancy of the phases.

As a first approximation,  let
$\{{\phi^{V}_{j}}^{\dagger},{\phi^{H}_{j}}^{\dagger} \}$ denote the phases
of the no-vortex solution for the {\it infinite} ladder, i.e., in the absence
of edge effects.  To ease the notation, let ${\phi^{H}}^{\dagger}$ denote the
common phase of the horizontal  junctions (so
${\phi^{H}_{j}}^{\dagger}={\phi^{H}}^{\dagger}$ for all $j$),  and define
${\phi^{V}}^{\dagger}$ similarly for the phases of the vertical junctions.
The only physically acceptable solution of
equations~(\ref{eq:scequation1})--(\ref{eq:scequation2}) that also satisfies
the stability condition~(\ref{eq:horizontalcond})  is:
\begin{equation}
\label{eq:scsolnoed}
{\phi^{V}}^{\dagger}=\arcsin I, \hspace{.2in} {\phi^{H}}^{\dagger}=\pi f,
\end{equation}
where $0 \leq f \leq 1/2$ and all the angles are restricted to
the first quadrant.

This solution exists if and only if $I \leq 1$.
A linear stability analysis
shows that, for all $I < 1$, the solution is stable
if ${\phi^{V}}^{\dagger} = \arcsin I \in [0,\pi/2)$.
The other possible solutions with
${\phi^{V}}^{\dagger} =  \pi - \arcsin I$
or ${\phi^{H}}^{\dagger} =  \pi f -\pi$ are linearly
unstable~\cite{mythesis}.
In summary, when the edges are completely neglected,
the array behaves like a single junction:
its only stable no-vortex solution of the observed
form~(\ref{eq:scsolnoed})
disappears at $I=1$ through a saddle-node bifurcation.
This existence criterion will be used in Section~\ref{sec:bifurc}
when discussing the stability properties of the nv solution.

Figure~\ref{fig:scnovortex} shows that this infinite-ladder approximation
works well near the center of the ladder, but
breaks down close to the edges. We now take edge effects into account by
considering an edge-corrected solution (denoted by $\ddagger$)
\begin{equation}
\label{eq:scnvedge}
{\phi^{V}_{j}}^{\ddagger}={\phi^{V}}^{\dagger}+A_{j}, \hspace{.5in}
{\phi^{H}_{j}}^{\ddagger}={\phi^{H}}^{\dagger}-B_{j}.
\end{equation}
where $\{A_{j}, B_{j}\}$ denote the corrections.  From the fixed point
equations (\ref{eq:scequation1}), (\ref{eq:scequation2}),
the $\{A_{j},B_{j}\}$ must satisfy
\begin{eqnarray}
\label{eq:corrections1}
&\hspace{-.5in} I+ \sin (\pi f - B_{j-1}) = \nonumber \\
& \hspace*{1in} \sin (\pi f- B_{j})+ \sin (\arcsin I + A_{j})\\
&A_{j}-A_{j+1}+2 B_{j}=0.
\label{eq:corrections2}
\end{eqnarray}
The corrections $\{A_{j},B_{j}\}$ are expected to be small,
except in a region very close to the edges. Thus,
Equation~(\ref{eq:corrections1}) can be expanded to first order in $A_{j}$
and $B_{j}$ to obtain a second-order difference equation for $A_{j}$:
\begin{eqnarray}
\label{eq:difference}
A_{j+1}-2 \alpha A_{j} + A_{j-1} =0\\
{\rm with}\;\; \alpha=1+\frac{\sqrt{1-I^{2}}}{\cos \pi f},
\label{eq:scnvalpha}
\end{eqnarray}
from which the horizontal corrections are
\begin{equation}
\label{eq:correctionhor}
B_{j}=\frac{A_{j+1}-A_{j}}{2}.
\end{equation}

The general solution of~(\ref{eq:difference}) is
\begin{eqnarray}
\label{eq:correctionvert}
A_{j}=P r^{j-(N+1)}+Q r^{1-j}\\
{\rm where} \; \; r=\alpha + \sqrt{\alpha^{2}-1} \equiv e^{1/\lambda}.
\label{eq:scnvr}
\end{eqnarray}
Hence, the edges produce corrections that
decay exponentially from both ends with a characteristic
length $\lambda(I,f)$. This $\lambda=1/\ln r$  is a measure of how
small perturbations decay inside of a region with the no-vortex
superconducting solution. A similar result was recently obtained by
Denniston and Tang~\cite{denniston} using the transfer matrix method for
the particular $I=0$ case.

To complete the solution, the constants $P, Q$ in~(\ref{eq:correctionvert})
have to be fixed from the boundary conditions
\begin{eqnarray}
\label{eq:scboundcond1}
 I = \sin(\pi f -B_{1}) + \sin(\arcsin I + A_{1}) \\
 I + \sin(\pi f -B_{N}) = \sin(\arcsin I + A_{N+1}),
\label{eq:scboundcond2}
\end{eqnarray}
which result from current conservation at nodes $1$ and $N+1$ respectively.
Since $\{A_{j},B_{j}\}$ become largest at the edges,
equations~(\ref{eq:scboundcond1})--(\ref{eq:scboundcond2}) are solved
numerically {\it without} linearization.
When the array is long enough, such that $\lambda \ll N+1$,
the effect of one edge on the other is negligible and
the solution is further simplified as
equations~(\ref{eq:scboundcond1})--(\ref{eq:scboundcond2}) decouple. Then
$Q$ and $P$ are obtained independently by solving
\begin{eqnarray}
\label{eq:scnvleft}
I= \sin\left(\pi f + \frac{Q}{2}\left(1-\frac{1}{r}\right)\right)
+ \sin(\arcsin I + Q)\\
I+\sin\left(\pi f-\frac{P}{2}\left(1-\frac{1}{r}\right)\right) =
\sin(\arcsin I +P).
\label{eq:scnvright}
\end{eqnarray}

Figure~\ref{fig:scnovortex} (a)--(b) shows that the above analytical solution
agrees well with the results of simulations for long ($N=25$) arrays.
The approximation accounts well for the effect of the open ends also for
short ($N=7$) arrays (not shown). The exponential decay of the perturbation
from the edges
is also checked satisfactorily in Figure~\ref{fig:scnovortex}(c).
As expected on physical grounds, the edge effects become more important
as both the field and the current are increased.
Thus, the approximation is best when $f$ and $I$ are
small, and worsens as  $I \rightarrow 1$  and $f \rightarrow 1/2$
($f=1/2$ is a singular limit,  as seen from the vanishing denominator
in~(\ref{eq:scnvalpha})).
This establishes limits on the use of this approximation
for the prediction of the depinning current at high values of the
frustration.

Our analytical approximation also explains other features of the numerics.
For instance, the corrections $A_{j}, B_{j}$ are spatially asymmetric
with respect to the center of the array when $I>0$
--- as seen in Figure~\ref{fig:scnovortex}(a)--(b) by
comparing the rightmost and leftmost phases. Note also that for $0< f \leq 1/2$
the largest vertical phase occurs at the right end of the ladder---
in obvious connection with the
preferred direction for flux-propagation in the array (${\bf -\hat x}$).
Moreover, it can readily be shown that
the change of the frustration from $f$ to $1-f$ has only one effect:
the vertical phases for frustration $1-f$ are a
mirror image, with respect to the center of the array,
of the vertical phases with frustration $f$. This implies that
the depinning current will be {\it identical} for both values of the
frustration, as expected, but the direction of propagation is
reversed~\cite{mythesis}.

\subsection{Single-vortex solution}
\begin{figure}[t]
\psfig{file=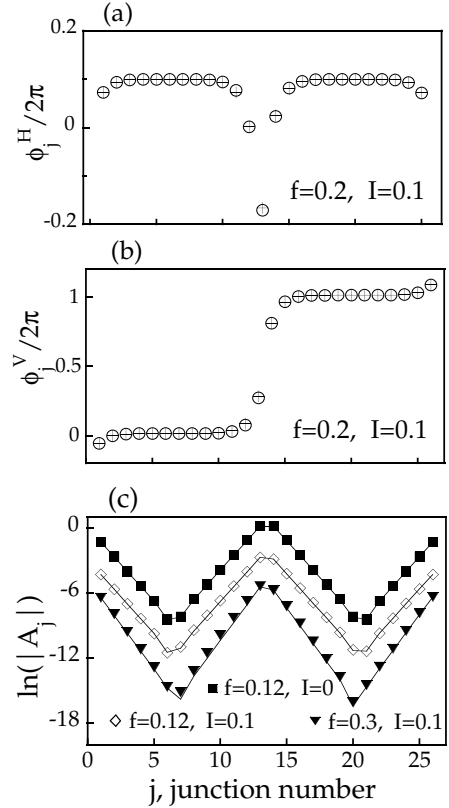,width=3.5in}
\vspace{.2in}
\caption{Single-vortex solution. (a) Phases of the horizontal junctions
for a $25 \times 1$ ladder for $I=0.1$ and $f=0.2$ with a vortex in the
central plaquette: numerical configuration from dynamical simulations (+), and
analytical approximation ($\circ$) as given
by~(\protect{\ref{eq:onevortex}}).
(b) Same as in (a) for the vertical junctions. The phases are not
reduced to the $[-\pi,\pi)$ interval.
(c) Exponential decay of the correction $A_{j}$ from both the edges and from
the center of the vortex for different $f$ and $I$.
Symbols represent numerical simulations and the
solid lines are the corresponding analytical predictions given
by~(\protect{\ref{eq:Aonevortex}}).
The different graphs have been offset for clarity.}
\label{fig:onevortex}
\end{figure}

An analytical approximation for the single-vortex configuration can be
obtained in a similar fashion by realizing that the effect of a vortex
located in cell $a$ of the array is similar to the edge effects in the
no-vortex state. Note how, if the phases in 
Figure~\ref{fig:onevortex}(a)--(b) were reduced to $[-\pi,\pi)$, 
the single-vortex configuration is
composed of two halves, each of which is equivalent to a no-vortex 
superconducting solution
when we move away from the edges and from the vortex center $a$.

Hence, the zeroth order single-vortex solution
$\{{\phi^{V}_{j}}^{\dagger},{\phi^{H}_{j}}^{\dagger}\}$
is identical to the no-edge nv solution given in~(\ref{eq:scsolnoed}).
And the edge and vortex-corrected approximation
with a vortex distribution $n_{a}=1$ and $n_{j}=0, \;\; \forall j \neq a$
is given by
\begin{equation}
\label{eq:onevortex}
{\phi^{V}_{j}}^{\ddagger}=\arcsin I+ A_{j}, \;\;
{\phi^{H}_{j}}^{\ddagger}= \pi f - B_{j},
\end{equation}
where the corrections $\{A_{j}, B_{j}\}$ result now both from
the presence of the edges and of the vortex in plaquette $a$.

Following the same steps as for the nv configuration, we obtain
identical expressions~(\ref{eq:corrections1})--(\ref{eq:scnvr}) for the
corrections for {\it each} half of the array:
\begin{eqnarray}
\label{eq:Aonevortex}
A_{j}&=&\left \{
\begin{array}{ll}
P r^{j-a}+Q r^{1-j},& j \leq a\\
P' r^{j-(N+1)}+Q' r^{a+1-j}, & j > a
\end{array} \right .\\
B_{j}&=&\frac{A_{j+1}-A_{j}}{2}, \hspace{1in} j \neq a
\label{eq:Bonevortex}
\end{eqnarray}
where $r$ is, once more, given by (\ref{eq:scnvr}).
Therefore, the single-vortex solution is obtained by {\it matching} two
edge-corrected no-vortex solutions. In fact, the vortex in cell
$a$ effectively introduces two new ``edges'', at $a$ and $a+1$,
which also produce similar exponentially decaying corrections.
The matching condition at $a$ and $a+1$ is given by
the fluxoid quantization condition in the cell containing the 
topological vortex:
$A_{a}-A_{a+1}+2B_{a}= 2 \pi$.  Thus,
\begin{equation}
\label{eq:Baonevortex}
B_{a}= \pi +\frac{P' r^{a-N}+Q'-P-Q r^{1-a}}{2}.
\end{equation}

This completes the equations needed to determine
the unknowns $P,Q,P',Q'$ in our solution~(\ref{eq:Aonevortex}).
They can be calculated numerically~\cite{mythesis}, for given $I$ and $f$,
from~(\ref{eq:Baonevortex}) and the boundary conditions
from current conservation at nodes $1,a, a+1$ and $N+1$.
The approximation is compared with numerical simulations in
Figure~\ref{fig:onevortex} with excellent agreement, especially at small $f$.

We also note that although the ladder equations~(\ref{eq:scequation1})-
(\ref{eq:scequation2}) can be reduced approximately to
a discrete sine-Gordon equation~\cite{kardar1,kardar2,strouddyn},
our analytical expression~(\ref{eq:onevortex}) is a better approximation
than the much-used sine-Gordon kink, which is a good description in
strictly 1-D parallel arrays~\cite{parallelcontinuum}. 
A detailed  comparison of both approaches is
presented in Appendix~\ref{sec:appsg}.

\subsection{Fully-frustrated solution}

\begin{figure}[t]
\psfig{file=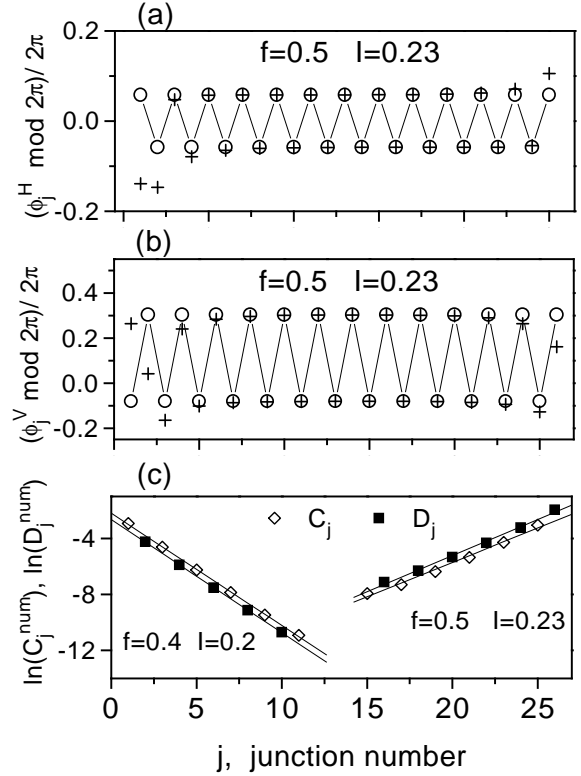,width=3.5in}
\caption{Fully-frustrated solution. (a) Phases of the horizontal junctions for
a $25 \times 1$ with $I=0.23$ and $f=0.5$. The numerical solution (+)
and infinite-ladder analytical approximation ${\phi^{V}_{j}}^{\dagger}$
(---$\circ$---)
are seen to coincide except close to the edges.
The solid line (a guide to the eye) emphasizes the wavelength equal to
two plaquettes.
(b) Same as (a) for the vertical junctions.
(c) Exponential decay of the corrections from the edges in the
fully-frustrated solution for varying $f$ and $I$.
Both the odd ($C^{\rm num}_{i}$) and even ($D^{\rm num}_{i}$) site corrections
have a characteristic length $\pm (\ln r)/2$ given
by~(\protect{\ref{eq:rschalfedge}}).
The solid lines are best linear fits with slopes $-0.805$ 
(for $f=0.4, I=0.2$) and
$0.502$ (for $f=0.5, I=0.23$). They are in  excellent agreement with
the predicted values $(\ln r)/2$ from~(\protect{\ref{eq:rschalfedge}}),
which are $0.806$  and $0.488$, respectively.}
\label{fig:schalfedge}
\end{figure}

The other relevant superconducting state is the fully-frustrated
solution, which appears in simulations when
$f \approx 1/2$ (Figure~\ref{fig:schalfedge}).
To obtain an analytical approximation, we follow once more the
same scheme as above: first, calculate a
no-edge basic solution; then, introduce the edges perturbatively.

In this case, the basic solution
is seen numerically to oscillate in space with a wavelength equal to two
plaquettes. Thus, when edges are neglected (the infinite-ladder
approximation), the phases can be approximated in general by
\begin{equation}
\label{eq:halfsolution}
\left \{
\begin{array}{l}
{\phi^{V}_{j}}^{\dagger}=2 \pi \; [a + (-1)^{j} b] \\
{\phi^{H}_{j}}^{\dagger}=2 \pi \; [c + (-1)^{j} d]
\end{array}
\right.
\end{equation}
where $a,b,c,d$ are to be determined from
(\ref{eq:scequation1})--(\ref{eq:scequation2}) with
$n_{j}=[1 \mp (-1)^{j}]/2$.
First, substitution in Eq.~(\ref{eq:scequation2}) gives $c$ and $d$
\begin{eqnarray}
c&=&f/2-1/4\\
d&=&b \pm 1/4.
\end{eqnarray}
Second, from Eq.~(\ref{eq:scequation2}) we obtain
\begin{eqnarray}
&&\sin 2 \pi a \; \cos 2 \pi b = I \nonumber\\
\sin 2 \pi b \; &&\cos 2 \pi a- 2 \sin \pi f \; \cos 2 \pi b=0, \nonumber
\end{eqnarray}
from which we then solve explicitly
for $a$ and $b$ in terms of the parameters $f$ and $I$
\begin{eqnarray}
\label{eq:halfsolution2}
a&=&\frac{1}{2 \pi} \arcsin \sqrt{L/2}\\
b&=&\frac{1}{2 \pi} \arccos \sqrt{2 I^{2}/L}
\end{eqnarray}
where
\begin{equation}
\label{eq:Lhf}
L=(1+I^{2})\pm \sqrt{(1-I^{2})^{2}-16 I^{2} \sin^{2}\pi f}.
\end{equation}
Figure~\ref{fig:schalfedge} (a)--(b)
compares the analytical infinite-ladder fully-frustrated
approximation~(\ref{eq:halfsolution})--(\ref{eq:Lhf}) with numerical
simulations. The agreement is good except near the ends, as expected.

The approximation above also yields
an existence criterion for the no-edge fully-frustrated solution.
In~(\ref{eq:Lhf}), the expression inside the square root must be
non-negative; hence
the infinite-ladder fully-frustrated solution does not exist if
\begin{equation}
\label{eq:currhalf}
I > I_{\rm ff,th} = \sqrt{4 \sin^{2} \pi f +1} - 2 \sin \pi f,
\end{equation}
where the subscript ``ff'' denotes fully-frustrated and ``th'' denotes a
theoretical approximation of a bifurcation condition.  We will use this
condition~(\ref{eq:currhalf}) later when we discuss the depinning of the
fully-frustrated solution.

This predicted form of the infinite-ladder fully-frustrated solution agrees
with previous findings obtained for the
special case when there is no driving current~\cite{benedict}. For $I=0$, our
solution~(\ref{eq:halfsolution})--(\ref{eq:Lhf})
 reduces to the stable configuration~\cite{benednote}
\[ {\phi^{V}_{j}}^{\dagger}=(-1)^{j} \arctan (2), \hspace{.2in}
{\phi^{H}_{j}}^{\dagger}=(-1)^{j+1} \arctan(1/2),\]
which coincides with the ground state for $f=1/2$ and $I=0$ calculated
by Benedict~\cite{benedict}. (To obtain this result from the expressions
above, note that $2I^{2}/L \rightarrow (1+4 \sin^{2} \pi f)^{-1}$ as $I
\rightarrow 0$, for the solution corresponding to the minus sign in
(\ref{eq:Lhf}).)
We also note that the infinite-ladder fully-frustrated solution
exists {\it for all} $f$ when $I=0$, i.e., there is no critical magnetic
field below which it ceases to exist, although it is energetically most
favorable when $f \approx 1/2$.

The physical meaning of this solution is clear: it contains a
topological vortex in every other cell, as seen  from the alternating sequence
of zeros and ones for the plaquette integers $\{n_{j}\}$ in the
equations~(\ref{eq:scequation1})--(\ref{eq:scequation2}).

In fact, although the solutions with
$\{n_{\rm odd}=1,\; n_{\rm even}=0\}$ and $\{n_{\rm odd}=0,\; n_{\rm even}=1\}$
are degenerate in an infinite array,
they are not so if the array is finite.
However, we will show in Section~\ref{sec:bifurc} that the depinning
of the fully-frustrated state is basically unaffected by the parity of the
number of cells in the ladder.

As we did for the no-vortex solution, we now introduce corrections
from the edges.  The improved solution
$\{{\phi^{V}_{j}}^{\ddagger},{\phi^{H}_{j}}^{\ddagger}\}$ is given by
\begin{equation}
\left \{
\begin{array}{l}
{\phi^{V}_{2i-1}}^{\ddagger}={\phi^{V}_{\rm odd}}^{\dagger}+C_{i}\\
{\phi^{V}_{2i}}^{\ddagger}={\phi^{V}_{\rm even}}^{\dagger}+D_{i}
\end{array}
\right. ,
\left \{
\begin{array}{l}
{\phi^{H}_{2i-1}}^{\ddagger}={\phi^{H}_{\rm odd}}^{\dagger}-E_{i}\\
{\phi^{H}_{2i}}^{\ddagger}={\phi^{H}_{\rm even}}^{\dagger}-F_{i}
\end{array}
\right.
\end{equation}
where $i=1,\ldots,\rm{ceil}(N/2)$ and there is
an additional ${\phi^{V}_{N+1}}^{\ddagger}$ when $N$ is even and
${\phi^{H}_{N+1}}^{\ddagger}=0$ when $N$ is odd.
The double cell is used to simplify the calculations, as suggested by the
spatial periodicity of the infinite-ladder solution.

Again, far away from the ends the corrections are small and we linearize the
governing equations (\ref{eq:scequation1})--(\ref{eq:scequation2})
around the basic solution (\ref{eq:halfsolution}). Thus
we obtain the following system of coupled difference equations:
\begin{eqnarray}
&&C_{i}-D_{i}+2E_{i}=0\\
&&D_{i}-C_{i+1}+2F_{i}=0\\
-E_{i} \cos {\phi^{H}_{\rm odd}}^{\dagger}&&=
-F_{i} \cos {\phi^{H}_{\rm even}}^{\dagger}
+D_{i} \cos {\phi^{V}_{\rm even}}^{\dagger}\\
-F_{i} \cos {\phi^{H}_{\rm even}}^{\dagger}&&=
-E_{i+1} \cos {\phi^{H}_{\rm odd}}^{\dagger}
+C_{i+1} \cos {\phi^{V}_{\rm odd}}^{\dagger}.
\end{eqnarray}
Eliminating $E_{i}, F_{i}$ and $D_{i}$ we get
a second-order difference equation for $C_{i}$:
\begin{equation}
C_{i+2}+2 \gamma C_{i+1} + C_{i}=0,
\end{equation}
with $\gamma(I,f)$ given by
\begin{equation}
\gamma=\frac{\sin^{2} \pi f + cb^{2} \cos 2 \pi f - 2 \left [
(\sin^{2} \pi f- sa^{2}) sb^{2} + ca^{2} \right
]}{\sin^{2}\pi f -cb^{2}}.
\end{equation}
(Here $sa, sb, ca$ and $cb$ are shorthand for
$\sin 2\pi a$, $\sin 2\pi b$, $\cos 2\pi a$ and $\cos 2\pi b$ respectively).
This difference equation has the general solution
\begin{eqnarray}
\label{eq:Cschalfedge}
&C_{i}= P r_{\rm ff}^{i} + Q r_{\rm ff}^{-i}\\
{\rm with} \;\; &r_{\rm ff}=-\gamma + \sqrt{ \gamma^{2}-1}
\equiv e^{1/\lambda_{\rm ff}}.
\label{eq:rschalfedge}
\end{eqnarray}
And $r_{\rm ff}$ is related again to another characteristic penetration
depth for the perturbations from the edges to die off, this time inside
a region with the fully-frustrated solution.
The coefficients $P$ and $Q$ have to be calculated numerically using the
boundary conditions from nodes $1$ and $N+1$.
The spatial dependence of $D_{i}$ is also of the same form
$D_{i}= R r_{\rm ff}^{i} + S r_{\rm ff}^{-i}$.
Note that in both equations, $i$ is the number that indexes the double cell.

Figure~\ref{fig:schalfedge} (c) illustrates the accuracy of these
approximate formulas. Specifically, we plot the spatial dependence of the
predicted deviations
\[ C_{i}^{\rm num}= \phi^{V}_{2i-1}- {\phi^{V}_{2i-1}}^{\dagger}, \hspace{.2in}
 D_{i}^{\rm num}= \phi^{V}_{2i}- {\phi^{V}_{2i}}^{\dagger},\] where
${\phi_{j}^{V}}$ is obtained from the numerical solution and
${\phi_{j}^{V}}^{\dagger}$ is the infinite-ladder approximation. The expected
exponential decay close to the edges with characteristic
length $\lambda_{\rm ff}=2/\ln r_{\rm ff}$ is verified in the figure.

\section{Dynamical depinning transitions of the array}
\label{sec:depinning}
In this section, we describe the depinning transitions as seen
in dynamical simulations of the ladder array at zero temperature.
In the following section, we will explain these dynamical results
by relating them to the bifurcations of the
no-vortex, single-vortex, and fully-frustrated superconducting solutions.
The main goal is to give a
rigorous mathematical explanation of the following observations:
Dynamical simulations show that
the array is superconducting at low values of the driving current $I$.
As $I$ is increased, the array remains superconducting until a critical current
$I_{\rm dep}(f)$ is reached, after which the array depins and develops a
non-zero average voltage. This depinning current
$I_{\rm dep}$  decreases monotonically as the frustration $f$ increases from
$0$ to  $1/2$.

These numerical observations are all at the averaged level of the $IV$
characteristics. They do not tell us anything about the detailed
configuration of the individual junctions.  In particular, there are
several distinct superconducting states (e.g., the nv, sv, and ff
states discussed in Section~\ref{sec:analytical}, and states
containing multiple vortices) but these are indistinguishable on the $IV$
curve.  This ambiguity raises the question: what is the state of the ladder
just before it depins?

We will show in Section~\ref{sec:bifurc} that for most values of $f$, the
depinning of the ladder is caused by the destruction of the no-vortex
state.  Specifically, the global
depinning current $I_{\rm dep}(f)$ can be predicted by calculating the
current at which the no-vortex state is annihilated in a saddle-node
bifurcation.  The only exception occurs for values of $f$ close to $1/2$,
where the depinning is due to saddle-node bifurcations of states of the
fully-frustrated type.  The noteworthy point here is that no other
superconducting states play a role in the global depinning of the array.

\begin{figure}[t]
\psfig{file=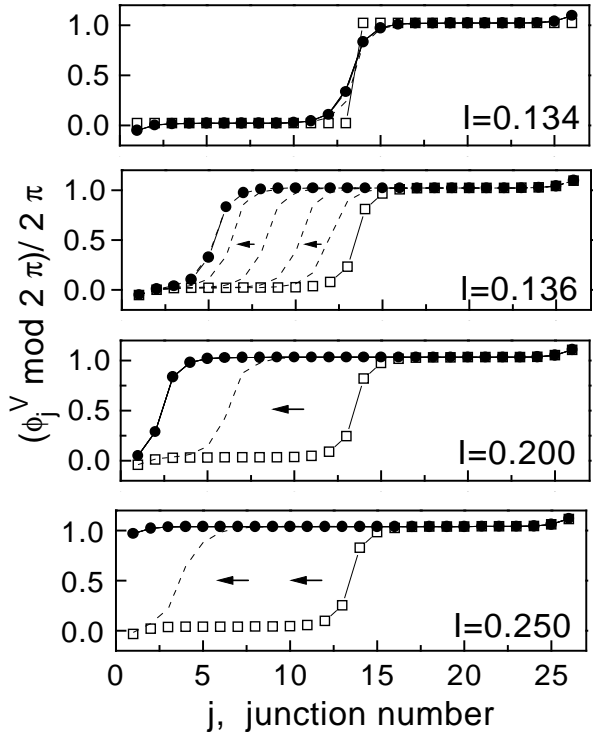,width=3.5in}
\caption{Snapshots of the time-evolution of the single-vortex
solution for a $25 \times 1$ ladder array with $\beta_{c}=10$ and $f=0.2$ and
increasing $I$. Initial ($\Box$), intermediate($--$), and final ($\bullet$)
configurations are shown.
At $I=0.134<I_{\rm LAT}$ the initial configuration
with a $2 \pi$-jump relaxes to the stationary single-vortex solution.
When $I=0.136 \simeq I_{\rm LAT}$ the vortex becomes dislodged from the center.
It moves slowly to the left and then stops at some intermediate position
between the center and the edge.
For $I_{\rm LAT}<I=0.2<I_{\rm left}$
the vortex moves until it gets pinned near the edge, where the potential
barrier is larger. Finally, at $I_{\rm left}<I=0.25<I_{\rm dep}$ the vortex
is expelled from the array and the {\it no-vortex solution} is recovered.}
\label{fig:LAT}
\end{figure}

However, the question arises as to how the depinning behavior would change 
when configurations with vortices are used as initial conditions (rather
than the random or zero-phase initial conditions that we ordinarily use in
our simulations of the $IV$ characteristics).
To address this issue, we perform dynamical simulations from an
initial condition with one $2 \pi$-step in the middle cell of the array:
\[\phi^{H}_{j}(t=0)=0, \; \; \phi^{V}_{j}(t=0)=
2 \pi \; \Theta\left(j-{\rm ceil}\left(\frac{N+1}{2}\right)\right),\]
where $\Theta(x)$ is the Heaviside step function.
This initial condition is {\it not} a solution of the system and, thus,
under the dynamical equations (\ref{eq:eq1})--(\ref{eq:eq3}) it relaxes
onto a true solution for the ladder.
For most $\{f,I\}$, the {\it single-vortex superconducting state}
(top panel of Figure~\ref{fig:LAT}) is reached, i.e.,
a static configuration with a topological vortex
in cell $a$ of the array such that $n_{a}=1$ and $n_{j}=0, \;
\forall j \neq a, \; j=1,\ldots,N$.
For some ranges of $f$ and $I$, this configuration
is not dynamically stable and other solutions are found, as discussed
in Section~\ref{sec:bifurc}.

The numerical observations shown in Figure~\ref{fig:LAT} depict the dynamical
behavior of the single-vortex state for most values of $f$, i.e.
approximately $0.12 < f < 0.37$. They can be summarized as follows:
For $f>f_{\rm min}$ and small driving current $I$, the system relaxes
onto a static single-vortex solution with a vortex in the middle of the array.
The solution is slightly distorted as the current is increased,
until at $I=I_{\rm LAT}(f)$ the vortex moves from the center cell toward
the left.
(This current is analogous to the well-known
Lobb-Abraham-Tinkham (LAT) depinning current
for two-dimensional Josephson junction arrays~\cite{LAT}).
For currents very close to $I_{\rm LAT}$, the vortex moves slowly
and gets trapped in another cell somewhere between the center and the edge.
For somewhat larger $I > I_{\rm LAT}$, the vortex moves all the way to the
left edge where
it becomes pinned until, at a second critical current
$I_{\rm left}$, it is
expelled from the array and the no-vortex solution is recovered again.
The no-vortex configuration then remains
stable until, at $I_{\rm dep}$,  global depinning of the array occurs.
If instead of placing the vortex in the middle, we locate it closest to the
{\it right} edge, it depins at a current $I_{\rm right}< I_{\rm LAT}$,
and moves toward the center.

These observations can be clarified with the usual analogy~\cite{terryvortex}
of the vortex as a damped particle moving in a sinusoidal potential under
the action of a $\bf{- \hat x}$ Lorentz-like force directly proportional to
$I$. The maxima of the potential correspond to the vertical junctions and
the minima are situated in the middle of the cells.
Thus, the initial barrier which has to be overcome to begin the motion
explains the critical $I_{\rm LAT}$.
Moreover, the open boundaries can be thought to produce an
exponentially-decaying envelope superimposed on the sinusoidal potential.
Thus the motion of the vortex is easier when the vortex is close to
the right edge and becomes increasingly difficult as the left
boundary is approached. When $I_{\rm left}$ is reached, the vortex is able to
overcome the edge barrier and is expelled from the array.
Then the no-vortex configuration is recovered and no new vortex enters
the ladder.

We mentioned before that the described behavior is observed for values of
the frustration contained between two limiting values.
First, there is a minimum frustration $f_{\rm min} \sim 0.12$
below which the single-vortex
solution never ensues from this initial condition; in fact, the system
settles on the no-vortex superconducting solution.
Second, for $f$ larger than $\sim 0.37$, the vortex is {\it not} expelled
from the array at the left edge before depinning.
Depinning occurs in that case when
vortices enter the array from the right edge. These observations will
be clarified in Section~\ref{sec:bifurc}.

We have also performed dynamical simulations for the multi-vortex case
and reached similar conclusions for most values of $f$.
In that case, the initial condition consists of
$N f$ equally spaced $2 \pi$-steps which are then allowed to evolve
dynamically. For $I=0$ and large enough $f$, the initial condition relaxes
onto the expected solution with $N f$ vortices in the array.
As the current is increased,
these move towards the left end where they accumulate until they
are expelled one by one at different currents. After this, the no-vortex
solution is again recovered.

On the other hand, the picture changes when $f$ is close to $1/2$. There,
solutions of the fully-frustrated type are obtained from the multi-vortex
initial condition and there is no expulsion of vortices from the ladder.
Instead, the array depins globally at the current where the
ff solution ceases to exist, i.e. $I_{\rm dep}(f \rightarrow 1/2)
\approx I_{\rm ff,th}$.

We conclude that in the ladder, the depinning of one vortex
(or a train of vortices) is {\it not} equivalent
to the global edge-dominated depinning of the device.
As we will confirm in Section~\ref{sec:bifurc},  the no-vortex and
fully-frustrated states are the relevant
solutions for the depinning of the array;
for moderate $f$, even if the initial conditions contain vortices,
these are expelled from the array as the current is increased and,
eventually, the no-vortex solution is recovered.
For $f$ close to $1/2$, the system settles onto fully-frustrated solutions
with distinct depinning properties.

We have also identified three other critical currents related
to the single-vortex configuration: $I_{\rm right}$, at which a vortex
at the right edge begins to move; $I_{\rm LAT}$, at which dynamical
depinning occurs for a single vortex centered in the middle of the ladder; and
$I_{\rm left}$, at which the vortex is expelled at the edge.

All of these dynamical observations are explained in detail in the next
section where they are compared to their exact mathematical descriptions.

\section{Bifurcation analysis of the depinning transitions}
\label{sec:bifurc}
In this section, we use bifurcation theory to obtain exact criteria
for all the critical currents of the no-vortex, single-vortex, and
fully-frustrated states. We have checked consistently that these bifurcations
explain the dynamical depinning behavior of the array as described in
Section~\ref{sec:depinning}.
Furthermore, analytical simplifications to some of
those criteria will be deduced from approximations of the
exact depinning results.

The depinning of the ladder can be explained in dynamical
terms as follows:
The linear stability of the superconducting states of the ladder
as a function of $I$
can be deduced from the Jacobian matrix ${\bf J_{\rm dyn}}$ of the 
dynamical system~(\ref{eq:eq1})--(\ref{eq:eq3}) for a given value of $f$,
and for a given fixed point---in particular, the no-vortex, single-vortex or
fully-frustrated state. 
If all the eigenvalues have negative real parts, the
fixed point is linearly stable.  As we increase $I$, some of the
eigenvalues move to the right in the complex plane, and the fixed point
becomes less stable.  The critical current for a given fixed point is
defined by the condition that the maximum of the real parts of the
eigenvalues becomes positive.  Then, to predict the {\it global} depinning
current, we compare the critical currents of the different superconducting
states, and take the maximum of those.  In other words, we predict that
global depinning occurs when the ``last'' stable state bifurcates.

Recall that there are several scenarios~\cite{steve}
by which a stable fixed point can undergo such a bifurcation. 
First, in a zero-eigenvalue bifurcation, a
single eigenvalue moves along the real axis, and passes from negative to
positive at the bifurcation.  There are three main subtypes of
zero-eigenvalue bifurcation: saddle-node, transcritical, and pitchfork.  In
the saddle-node bifurcation, a stable fixed point collides with a saddle
point, and both are annihilated.  In contrast, in the transcritical and
pitchfork bifurcations, the stable fixed point is not destroyed --- it
continues to exist but goes unstable.  A second scenario is provided by the
Hopf bifurcation which involves a pair of complex conjugate eigenvalues passing
through the imaginary axis from the left half plane to the right half plane
-- again this bifurcation destabilizes the fixed point, but does not
destroy it.

\begin{figure}[t]
\psfig{file=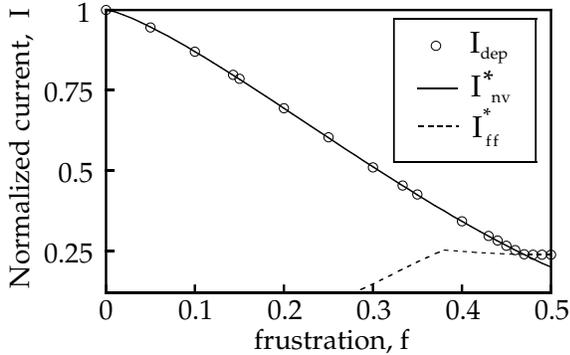,width=3.5in}
\caption{$f$-dependence of the global depinning current of the array. 
The numerical $I_{\rm dep}(\circ)$ is obtained by sweeping the current 
up from zero in dynamical simulations; no assumptions are made about 
the state of the system.
The static current $I_{\rm nv}^{\star}$ marks the point where a particular
superconducting state, the no-vortex solution, is destroyed in a
saddle-node bifurcation. Similarly,
$I^{\star}_{\rm ff}$ corresponds to the maximum of the saddle-node
bifurcation currents for solutions of the fully-frustrated type.}
\label{fig:depinning}
\end{figure}

Since it can be shown that Hopf bifurcations are not possible in this 
system~\cite{newpaper}, we can simplify our calculations by
using the {\it static} 
system~(\ref{eq:scequation1})--(\ref{eq:scequation2})
to identify the location of the zero-eigenvalue bifurcations.
Those bifurcation points are characterized by a change in the number of 
fixed points and, thus,   
from the implicit function theorem~\cite{calculus},
the Jacobian matrix ${\bf J}$ of the static system has zero determinant there.
Hence, we use the superconducting (static) up-down symmetric
system given by ${\bf f(x)= 0}$,
\begin{eqnarray}
\label{eq:systemeq}
&{\rm f}_{i}({\bf x})= I + \sin {\rm x}_{N+i} -\sin {\rm x}_{N+1+i} -&
\sin {\rm x}_{i}, \nonumber\\
&&i=1,\ldots,N+1\\
&{\rm f}_{N+1+i}({\bf x})={\rm x}_{i}-{\rm x}_{i+1}-
2{\rm x}_{N+1+i}+&2\pi(f-n_{i}), \nonumber\\
&& i=1,\ldots,N
\end{eqnarray}
with ${\bf x}=({\rm x}_{1},\ldots,{\rm x}_{2N+1})\equiv
(\phi^{V}_{1},\ldots,\phi^{V}_{N+1},\phi^{H}_{1},\ldots,\phi^{H}_{N})$.
For a given $f$ and a given superconducting state,
we compute the bifurcating fixed point $\bf x^{\star}$ and its associated
critical current  $I^{\star}(f)$ at which  ${\rm det({\bf J})}=0$.
To this end, we define an augmented algebraic system
with the current $I$ as an extra
variable, and the constraint on the determinant
as an extra equation. Then $\bf x^{\star}$ and
$I^{\star}$ are obtained by solving
${\bf F(X^{\star})= 0}$ where ${\bf X}=({\bf x},I)$ and
\begin{eqnarray}
\label{eq:augmented1}
{\rm F}_{j}({\bf X})=&&{\rm f}_{j}, \hspace{.2in}\;\; j=1,\ldots,2N+1\\
{\rm F}_{2N+2}({\bf X})=&&{\rm det({\bf J})}.
\label{eq:augmented2}
\end{eqnarray}
Figure~\ref{fig:depinning} shows that the dynamical depinning of
the ladder is explained by $I^{\star}$ (zero-eigenvalue) bifurcations.

The rest of this section is devoted to analyzing
the bifurcations of the no-vortex, single-vortex and fully-frustrated
configurations. For the sake of clarity, we follow a parallel scheme for all of
them and keep the notation consistent. For each of the states, we first
calculate numerically the zero-eigenvalue bifurcations
from~(\ref{eq:augmented1})--(\ref{eq:augmented2}) (always denoted with
a star $^{\star}$)
and compare them with the depinning currents from {\it dynamical} simulations
(always denoted with symbols in the figures). Then, when possible, we deduce
analytical simplifications of these criteria in one of two ways: (a) by
deducing that the instability is essentially caused by a bifurcation of the
no-edge solution, or (b) by explaining the instability as edge-originated
and, thus, localized at the boundaries.
These theoretical analytical approximations are always denoted with
the subscript ``th''.
Moreover, to emphasize the importance of
the edges in the depinning transitions, we calculate the depinning of
the no-edge solutions for all three configurations.
Finally, note that no energy arguments are invoked in this discussion.
Thermodynamic considerations are studied in detail
in Section~\ref{sec:conclusion} where
phase diagrams of these superconducting states for the
no-edge and finite ladders are presented.

\subsection{Bifurcation of the no-vortex solution}

Figure~\ref{fig:depinning} shows that for most values of $f$, the
depinning current $I_{\rm dep}(f)$ (obtained from dynamical simulations)
coincides with the critical current
$I_{\rm nv}^{\star}$ for the  no-vortex state
---calculated from the augmented
system~(\ref{eq:augmented1})--(\ref{eq:augmented2}).
Indeed, we find that the bifurcating phase configuration 
${\bf x}_{\rm nv}^{\star}$ matches the depinning 
configuration ${\bf x}_{\rm dep}$
observed in dynamical simulations. Hence, the bifurcation of the no-vortex
state constitutes an exact criterion for the global depinning current,
except for values of $f$ close to $1/2$,  where the global depinning is
caused by the destruction of the fully-frustrated solution, as explained
below.

\begin{figure}[t]
\psfig{file=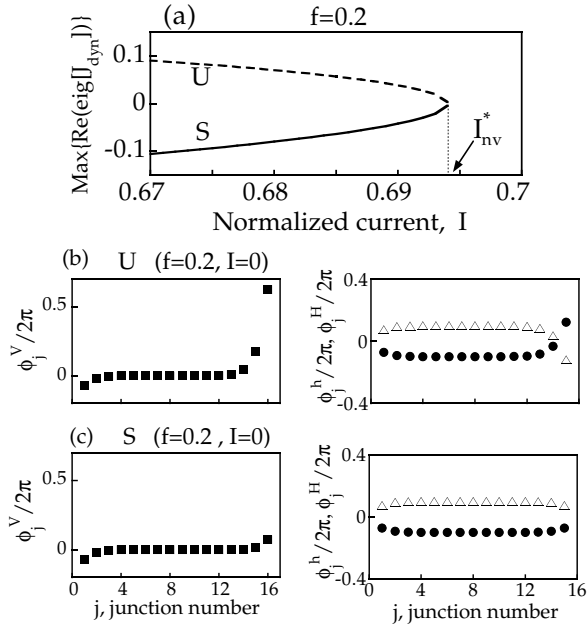,width=3.5in}
\caption{Saddle-node bifurcation of the no-vortex solution in a $15 \times 1$
ladder. (a) Value of the maximum of the real parts of the eigenvalues 
of the dynamic
Jacobian matrix ${\bf J_{\rm dyn}}$ for the stable (S, solid line) and unstable
(U, dashed line) branches with $f=0.2$ and $\beta_{c} =0.25$. They collide and
annihilate at $I^{\star}_{\rm nv}$ in a saddle-node bifurcation.
We remark that, although the eigenvalues change with $\beta_c$,
the bifurcation point $I^{\star}_{\rm nv}$ is independent of $\beta_{c}$.
(b) Phase configuration of the vertical (left) and horizontal
(right) junctions for the unstable branch (U) at $f=0.2$ and $I=0$.
(c) Same as (b) for the stable branch (S).
Note that both configurations (b)--(c) are up-down symmetric,
and the unstable branch (b) corresponds to an inverted pendulum at the
rightmost cell.}
\label{fig:sadnodnv}
\end{figure}

To gain intuition about how to derive analytical approximations for
$I_{\rm dep}(f)$, it is helpful to characterize the depinning bifurcation
more precisely.
Our numerical computations indicate that the depinning of the no-vortex
state is due to a saddle-node bifurcation.  As $I$ approaches
$I_{\rm nv}^{\star}$
from below, the stable no-vortex state approaches an unstable no-vortex
state, and coalesces with it when $I = I_{\rm nv}^{\star}$, causing both
states to
disappear.  Figure~\ref{fig:sadnodnv}(a) shows the maximum {\it dynamical}
eigenvalue for both
the stable and unstable states -- note that both of these eigenvalues are
pure real,
and they equal 0 at the critical current.  As expected, this  plot has the
standard shape of a saddle-node bifurcation diagram~\cite{steve}.
Figure~\ref{fig:sadnodnv}(b)--(c)
plots the phase configuration for both states at $I=0$.  They have
similar spatial structure, except near the rightmost cell, where (in the
language of the mechanical analog) the unstable state has an inverted
pendulum.

Incidentally, Figure~\ref{fig:sadnodnv} also shows that both states satisfy
the up-down symmetry
$\phi^{H}_{j}=-\phi^{h}_{j}$ discussed in Section~\ref{sec:system}.
Numerical simulations show that this symmetry continues to hold for all
values of $I$ on both the stable and unstable branches.  Thus, the global
depinning bifurcation takes place entirely within the invariant manifold of
up-down symmetric states -- it is {\it not} a symmetry-breaking
bifurcation.

\subsubsection{Analytical approximations for $I_{\rm nv}^{\star}(f)$}

The conclusion that the depinning transition for most values of $f$
corresponds to a saddle-node bifurcation of the no-vortex
superconducting state can be 
simplified further. We now obtain analytical approximations for
$I_{\rm nv}^{\star}(f)$ using the approximate solutions calculated in
Section~\ref{sec:analytical}.

We recall that the bifurcation of the infinite-ladder
no-vortex configuration does not explain the observed $f$-dependence of
the finite-ladder nv depinning. As discussed in Section~\ref{sec:analytical},
if the edges are neglected completely,
the no-vortex solution~(\ref{eq:scnvedge})--(\ref{eq:correctionvert})
is predicted to exist and be stable for all $I< 1$, independent of the
frustration $f$,
in analogy with the single junction. Thus, the depinning for the no-edge nv
state occurs through a saddle-node bifurcation
at $I^{\dagger}_{\rm nv,th}(f) =1, \;\; \forall f$.

To capture the observed $f$-dependence of the critical current
$I_{\rm nv}^{\star}$, we need a more careful approximation.  We now
present two such approximations to $I_{\rm nv}^{\star}$ that clarify
the physical picture of the transition.

The first strategy is to use the improved approximation~(\ref{eq:scnvedge})
for the  no-vortex state (in which the edge effects are included
perturbatively), and then write down a {\it simplified augmented system}
${\bf F^{\ddagger}(X^{\ddagger})=0}$ for this solution, similar to the
expression~(\ref{eq:augmented1})--(\ref{eq:augmented2})
for the full $(2N+2)$-dimensional system.
We can then calculate the critical current $I_{\rm nv,red}$ for this reduced
model, defined as the value of $I$ where the
{\it perturbative} solution undergoes a saddle-node bifurcation. This
renders the
calculation analytically tractable since, for a given $f$,
only three variables $\{P,Q,I\}$ suffice to describe the perturbative solution
(\ref{eq:scnvedge})---instead of $I$ and $2N+1$ phases for the full solution.
The simplified augmented system is constituted by
(\ref{eq:scnvleft}), (\ref{eq:scnvright}) together with the condition that
the determinant of the $2 \times 2$ Jacobian matrix equals zero,
corresponding to
the zero-eigenvalue condition at a saddle-node bifurcation.

Furthermore, since equations (\ref{eq:scnvleft}) and (\ref{eq:scnvright})
are uncoupled when $N$ is not very small, 
even this three-dimensional system can be further
reduced to a two-dimensional system with unknowns $P$ and $I$:
\begin{eqnarray}
\label{eq:reduced1}
{\rm F}^{\ddagger}_{1}&=&I +
\sin \left ( \pi f- \frac{P}{2} \left(1-\frac{1}{r}\right)\right)-
\sin (\arcsin I + P)\\
{\rm F}^{\ddagger}_{2}&=&
\frac{\partial {\rm F}^{\ddagger}_{1}}{\partial P}
\label{eq:reduced2}
\end{eqnarray}
where $r=r(f,I)$ is given by (\ref{eq:scnvr}).
Note that this is a set of {\it local} equations referred to the
rightmost end of the array.
We numerically solve the $2 \times 2$ reduced system
${\bf F}^{\ddagger}(P_{\rm red},I_{\rm nv,red})={\bf 0}$ to obtain the
approximate depinning current $I_{\rm nv,red}(f)$ and the value of the
rightmost phase
$P_{\rm red}$ at the bifurcation.

\begin{figure}[t]
\psfig{file=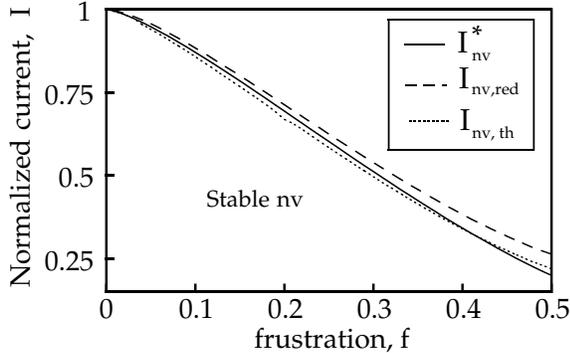,width=3.5in}
\caption{
Analytical simplifications $I_{\rm nv,red}$ and $I_{\rm nv,th}$
provide approximations to the critical current for the no-vortex
solution $I^{\star}_{\rm nv}$
by concentrating on the rightmost cell of the array.}
\label{fig:depanalyt}
\end{figure}

Figure~\ref{fig:depanalyt} shows that $I_{\rm nv,red}(f)$ predicts the exact
$I^{\star}_{\rm nv}(f)$ reasonably well. As expected, the prediction gets worse
as $f$ nears $1/2$ since the perturbative approximation of the no-vortex
solution is less accurate in that limit.

Both the analyses of the eigenfunctions of the full $2N+2$
system~(\ref{eq:augmented1})--(\ref{eq:augmented2})
and of the reduced system~(\ref{eq:reduced1})--(\ref{eq:reduced2})
indicate that the global depinning of the ladder is caused by a local
instability of the rightmost junction of the array.
This is consistent with physical arguments which imply that after depinning
occurs, vortices propagate in the array in the ${\bf - \hat x}$ direction
under the effect of a magnetic Magnus (Lorentz-like) force.

The key role played by the rightmost junction suggests a second simplification,
which we call a heuristic criterion for depinning. This criterion connects
the global depinning of the ladder with the much simpler
depinning transition in a single junction.  Recall that when the phase of a
single junction reaches  $\pi/2$, its superconducting solution is destroyed
in a saddle-node bifurcation~\cite{steve}.  Therefore, we intuitively
propose that when the phase of the rightmost
junction reaches $\pi/2$, the ladder depins.  Replacing the no-vortex
solution by its perturbative
approximation (\ref{eq:scnvedge}), we solve for the current $I_{\rm nv,th}$
by imposing
\begin{equation}
\label{eq:heurcrit}
{\phi^{V}_{N+1}}^{\ddagger}= \pi/2
\end{equation}
which implies
\begin{equation}
\arcsin I_{\rm nv,th} + P(I_{\rm nv,th})=\pi/2.
\end{equation}
Then, from (\ref{eq:scnvright}), we obtain
an implicit transcendental equation for $I_{\rm nv,th}(f)$:
\begin{equation}
\label{eq:heuristic}
\arcsin(1-I_{\rm nv,th})+\frac{r-1}{2r}\arccos I_{\rm nv,th}=\pi f
\end{equation}
with $r=r(I_{\rm nv,th})$ given by (\ref{eq:scnvr}).
This simple analytical prediction is shown to be in very good agreement
with the exact results in Figure~\ref{fig:depanalyt}.

The techniques described in this Section can be extended to include
the effect of self-fields. In Appendix~\ref{sec:appendixa}
we illustrate this approach and show how self-inductance modifies
the approximate no-vortex solution and the corresponding depinning current.

\subsection{Bifurcations of the single-vortex solution}

We begin the study of the stability of the single-vortex (sv) configuration by
considering the single-vortex far from the edges.
The results obtained for the vortex in the center will be used subsequently
to describe the effects of the edges on the sv state.

\subsubsection{Saddle-node bifurcation of the sv solution}
As described in Section~\ref{sec:depinning},
a vortex in the center of the array moves to the left over the potential
barrier when the critical current $I_{\rm LAT}$ is reached.
We show now that this depinning of the vortex
corresponds to a saddle-node bifurcation of the single-vortex solution.
To verify this, we restrict our attention to the single-vortex solutions
that are centered in the middle of the ladder; then we look
for the current $I^{\star}_{\rm sv}$ at which the determinant of
the static Jacobian matrix
${\rm det({\bf J}|_{{\bf x}^{\star}_{\rm sv}})} =0$.
As in (\ref{eq:augmented1})--(\ref{eq:augmented2}), we solve the augmented
system ${\bf F}({\bf x^{\star}_{\rm sv}},I^{\star}_{\rm sv}) = {\bf 0}$
to find where the centered single-vortex state ceases to exist.

\begin{figure}[t]
\psfig{file=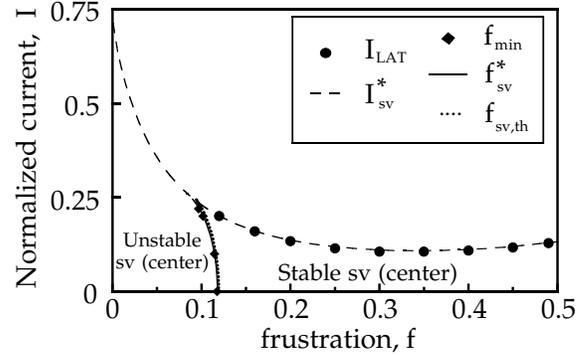,width=3.5in}
\caption{Stability diagram for the single vortex in the center of the
array.  $I_{\rm LAT} (\bullet)$ is calculated
dynamically from numerical simulations by sweeping  the current until a
vortex placed in the center of a $25 \times 1$ array moves.
These dynamical results $(\bullet)$ are well predicted by
the static $I^{\star}_{\rm sv}$ at which the fixed point corresponding
to the pinned vortex ceases to exist (dashed line).
Another dynamic instability of the single-vortex configuration at 
$f_{\min}$ (solid diamonds) is
identified as a symmetry-breaking subcritical pitchfork bifurcation
$f^{\star}_{\rm sv}$. It can be approximated with an
analytical criterion $f_{\rm sv,th}$ given
by~(\protect{\ref{eq:stabhorizontal}}).  This approximation is so
accurate that the curves for $f^{\star}_{\rm sv}$ and $f_{\rm sv,th}$ are
practically indistinguishable.}
\label{fig:depinvortex}
\end{figure}

Figure~\ref{fig:depinvortex} shows the perfect agreement between
the $I^{\star}_{\rm sv}(f)$ computed from the static augmented system
and the $I_{\rm LAT}(f)$ obtained from simulations where a vortex
is placed in the middle of the array and the current is increased until
it moves, i.e., $I_{\rm LAT}=I^{\star}_{\rm sv}$.

\begin{figure}[t]
\psfig{file=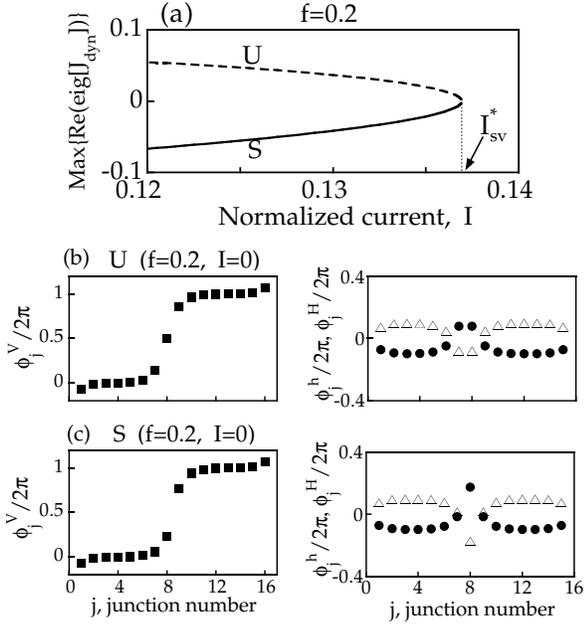,width=3.5in}
\caption{Saddle-node bifurcation of the single-vortex solution at the
center of a $15 \times 1$ ladder. (a) Value of
the maximum of the real parts of the eigenvalues of ${\bf J_{\rm dyn}}$ for
the stable (S, solid line) and unstable (U, dashed line) branches
with $f=0.2$ and $\beta_c = 0.25$. $I^{\star}_{\rm sv}$
is the point where a saddle-node bifurcation occurs for this particular
configuration. As in Fig.~\protect{\ref{fig:sadnodnv}}, the bifurcation
point is independent of $\beta_c$.
(b) Phases of the vertical (left) and horizontal
(right) junctions for the unstable branch (U) at $f=0.2$ and $I=0$.
(c) Same as (b) for the stable branch (S).
Again, both configurations (b)--(c) are up-down symmetric. Note that the
stable branch (c) is associated with a vortex located at the center of
a plaquette (a minimum of the potential), whereas the unstable branch (b)
corresponds to a vortex centered around a junction (a local maximum of
the potential energy).}
\label{fig:sadnodsv}
\end{figure}

Moreover,
Figure~\ref{fig:sadnodsv} confirms that this depinning transition is
indeed caused by a
saddle-node bifurcation.  An unstable single-vortex state collides with and
annihilates the stable single-vortex state at the transition.
Figure~\ref{fig:sadnodsv} shows that the two states have similar
spatial structure ---the
difference is that the stable state has its vortex in the center of a cell
(where the vortex sits in a potential well), while the unstable state has
its vortex on a junction (poised on a potential hill).

Although conceptually similar, our $I_{\rm LAT}$ for the ladder is not
equivalent to that calculated by Lobb-Abraham-Tinkham~\cite{LAT}. Their
current is  estimated by a static calculation of the energy barrier $E_{b}$
in an infinitely
extended two-dimensional array, while ours is the dynamic
current at which the centered single-vortex state undergoes a saddle-node
bifurcation in
the quasi-one-dimensional ladder. Moreover,
their static calculation does not include the effects of the
field $f$ or the injected current $I$ on the solutions while, in our
case, they are implicitly taken into account since the
configurations---and therefore their stability---depend
parametrically on $\{I,f\}$.

\subsubsection{Symmetry-breaking bifurcation of the sv solution}
We noted above the numerical observation~\cite{stroudscreen}
that, when performing dynamical simulations, the static single-vortex solution
is unstable below a {\it critical field} $f_{\rm min}(I)$.
We show now that this is the result of a symmetry-breaking instability
which is mathematically related to another zero-eigenvalue bifurcation.
Therefore, once again, the dynamical $f_{\rm min}$ coincides with a static
$f^{\star}_{\rm sv}$ calculated from the augmented
system~(\ref{eq:augmented1})--(\ref{eq:augmented2}) as the value of
$f$ where the determinant of the static Jacobian matrix {\bf J} is zero and, 
thus, a change in the number of fixed points is likely. 
Excellent agreement between
$f_{\rm min}$ and $f^{\star}_{\rm sv}$ is shown in
Figure~\ref{fig:depinvortex}.

\begin{figure}[t]
\psfig{file=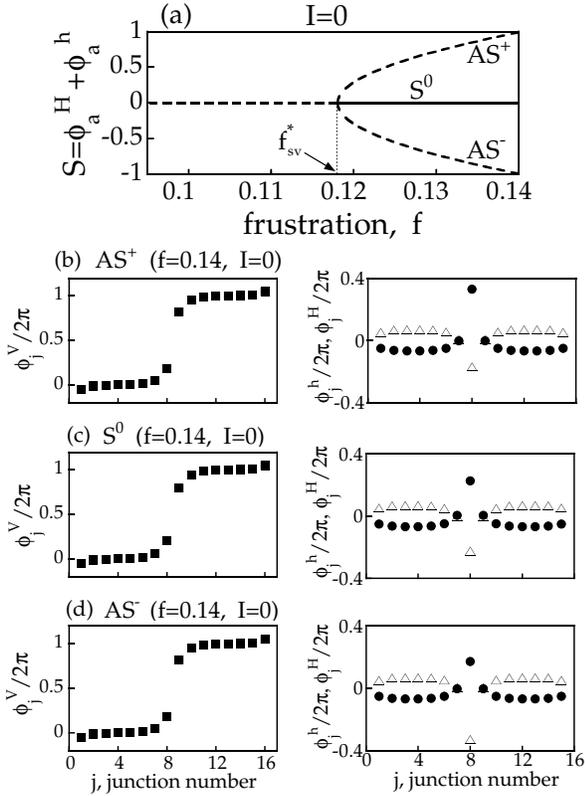,width=3.5in}
\caption{Subcritical pitchfork bifurcation of the single-vortex solution at
$f^{\star}_{\rm sv}$.
(a) Value of the up-down asymmetry $S= \phi^H_{a}+\phi^{h}_{a}$ for the
two asymmetric (AS$^{+}$, AS$^{-}$) and symmetric (S$^{0}$)
branches with $I=0$. The calculation of the dynamical eigenvalues (not shown)
indicates that the stable symmetric branch  becomes unstable at
$f^{\star}_{\rm sv}$ when it collides with the two unstable asymmetric
branches. For $f < f^{\star}_{\rm sv}$, only the unstable symmetric
branch survives.
(b) Phases of the vertical (left) and horizontal
(right) junctions for the asymmetric branch (AS$^{+}$) at $f=0.14$ and $I=0$.
(c) and (d) are the same as (b) but for the S$^{0}$ and the AS$^{-}$
branches respectively.
Note that although all three vertical configurations on the left panels 
look very similar, the (b) and (d) AS horizontal configurations on the right  
are up-down {\it asymmetric}.}
\label{fig:pitchforksv}
\end{figure}

Figure~\ref{fig:pitchforksv} depicts detailed information about this
bifurcation. Specifically, it shows three single-vortex states 
that co-exist
for $f$ slightly greater than $f^{\star}_{\rm sv}$.  These states appear
very similar,
but on close inspection, one notices that two of the states are asymmetric:
$\phi^{H}_j \neq -\phi^{h}_j$ (this is especially clear for the central
plaquette $j=a$.).  As $f \rightarrow f^{\star}_{\rm sv}$ from above, these
asymmetric states -- which are unstable -- simultaneously collide with the
stable symmetric state, rendering it unstable.  To visualize this
transition in greater detail, Figure~\ref{fig:pitchforksv}
plots the asymmetry for the
central plaquette, $S \equiv \phi^{H}_a + \phi^{h}_a$ , as a function of the
frustration $f$.  The symmetric state exists both above and below the
bifurcation, and satisfies $S=0$ throughout.  The two unstable branches
join the symmetric branch at $f=f^{\star}_{\rm sv}$.  The scenario depicted in
Figure~\ref{fig:pitchforksv}, common in symmetric systems, is known as a
subcritical pitchfork bifurcation~\cite{steve}.

 From this, we conclude that there exists a region of the $(f,I)$ plane
(Figure~\ref{fig:depinvortex})
where the single-vortex configuration exists but is always dynamically
unstable. In this region, the vortex (magnetic flux) is expelled
{\it transversally} from the array through
transient modes which do {\it not} preserve the up-down symmetry of the
horizontal phases.

We can also derive an analytical expression $f_{\rm sv,th}$
for the critical field $f^{\star}_{\rm sv}$.
As given in~(\ref{eq:horizontalcond}), all up-down symmetric
fixed points become unstable when
the phase of any of the horizontal junctions is larger than $\pi/2$ in
absolute value.
Since this largest phase occurs at the central plaquette (as seen in
Figure~\ref{fig:pitchforksv}),
the following stability criterion ensues from the condition
$\phi^{H}_{a} = -\pi /2$:
\begin{equation}
\label{eq:stabhorizontal}
2 \pi f_{\rm sv,th} + [P(f_{\rm sv,th}) - Q'(f_{\rm sv,th})] = \pi,
\end{equation}
where we have used~(\ref{eq:onevortex}) and (\ref{eq:Baonevortex}) and we
consider a long array such that the effect of the edges on the middle
cell can be neglected. Here $P, Q'$ have to be calculated from the
boundary conditions (current conservation) at nodes $a$ and $a+1$:
\begin{eqnarray}
I+ \sin(\pi f - &&P (r-1)/(2r))= \nonumber \\
  \sin( \arcsin I &&+ P) - \sin(\pi f - (Q'-P)/2) \nonumber\\
I - \sin(\pi f - &&(Q'-P)/2)= \nonumber \\
  \sin( \arcsin I &&+ Q') +\sin (\pi f + Q'(r-1)/(2r))
\label{eq:boundhorizontal}
\end{eqnarray}
with $r$ given by~(\ref{eq:scnvr}).
Numerical solution of the critical condition~(\ref{eq:stabhorizontal})
yields the curve $f_{\rm sv,th}(I)$ which is almost indistinguishable from
the curve for $f^{\star}_{\rm sv}$ from
the full augmented system (Figure~\ref{fig:depinvortex}).

It is especially interesting to check the case $I=0$.  Then $P=-Q'$ and
the critical condition~(\ref{eq:stabhorizontal})--(\ref{eq:boundhorizontal})
simplifies to
\begin{equation}
\label{eq:crithorizontal}
\sin \left[\pi f_{\rm sv,th}^{o} + \left(f_{\rm sv,th}^{o}-\frac{1}{2}\right)
\frac{\pi (r-1)}{2r} \right]
= \cos (\pi f_{\rm sv,th}^{o}) -1,
\end{equation}
which can be solved numerically to give
$f_{\rm sv,th}^{o} \equiv f_{\rm sv,th}(I=0) = 0.1193$, in agreement
with numerical findings from previous dynamical
simulations~\cite{stroudscreen}.

\subsubsection{Single-vortex configurations at the edges}

So far we have concentrated on a particular single-vortex state, namely one
where the vortex occurs in the middle cell of the array.  But there are
many other single-vortex states, each differing from the previous one by
displacing the vortex by one cell to the right or to the left.
Each of those configurations becomes unstable through bifurcations similar
to those discussed above for the case of the  centered single-vortex state.
For most values of $f$,
as discussed in Section~\ref{sec:depinning}, when the driving current $I$
is increased, a vortex in the ladder moves to the left (getting pinned
in cells closer to the left edge as $I$ grows) until it is expelled from
the array at the left boundary.
Thus, in explaining the effect of the edges on the sv solution
we are most interested in two critical currents:
the critical current $I_{\rm right}$, at which a vortex at the
right edge begins to move, and  the critical current
$I_{\rm left}$, at which the vortex is expelled at the left edge of
the ladder.

\begin{figure}[t]
\hspace*{-.5in}
\psfig{file=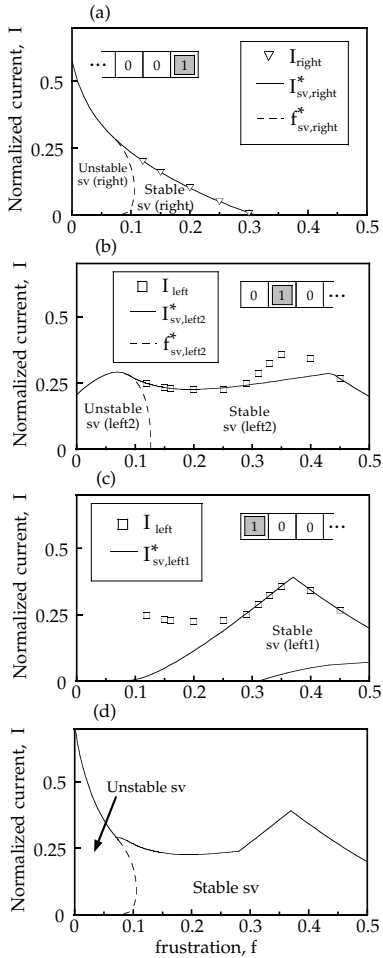,width=4in}
\caption{(a)--(c) Stability diagrams for the sv configuration when the vortex
is placed close to either edge. The symbols denote critical currents measured
through dynamical simulations. The solid lines ($I^{\star}_{\rm sv}$)
correspond to saddle-node bifurcations 
and the dashed lines ($f^{\star}_{\rm sv}$)
to subcritical pitchfork bifurcations where the flux is expelled transversally.
For instance, from (a) the vortex in the rightmost cell moves to the
left for $I>I^{\star}_{\rm sv,right}$. And $I_{\rm left}$ corresponds
to the expulsion of the vortex at the left edge for $f<0.37$.
In the region delimited by $I_{\rm right}$ and $I_{\rm left}$, there
is a cascade of saddle-node bifurcations which leads to the behavior
observed in Figure~\protect{\ref{fig:LAT}}.
For the top three
panels: (a) corresponds to the vortex in the rightmost cell; (b) in the second
cell; and (c) in the first (leftmost) cell.
(d) shows the region where at least one sv configuration (anywhere in the
ladder) is stable. It is the union of panels (a)--(c) and
Figure~\protect{\ref{fig:depinvortex}}.
}
\label{fig:svedge}
\end{figure}

To predict $I_{\rm right}$, we analyze the sv configuration
with the vortex placed in the rightmost cell (the $000 \ldots 01$
configuration). The results of the
analysis are similar to those for the vortex in the center. As shown in
Figure~\ref{fig:svedge}(a), this state can cease to exist through a
saddle-node bifurcation ($I^{\star}_{\rm sv,right}$), or become unstable
through a symmetry-breaking pitchfork bifurcation ($f^{\star}_{\rm sv,right}$).
The agreement with the dynamical simulations is excellent.

The rigorous explanation of $I_{\rm left}$ turns out to be slightly more
complicated. A careful examination of the numerics reveals that,
depending on the value of $f$, the vortex
can be expelled from the array in one of two ways: from the leftmost cell
(cell number 1) as expected, or directly from cell number 2. Thus, we need
to examine the dynamical stability of two sv states: the one with
the vortex in cell 1 ($1000 \ldots 00$) and in cell 2 ($0100 \ldots 00$).

The dynamic stability of the $0100 \ldots 00$ configuration contains no
new elements. The two observed bifurcations
(a saddle-node $I^{\star}_{\rm sv,left2}$ and
a subcritical pitchfork $f^{\star}_{\rm sv,left2}$) are similar to those
explained above. The results are presented in
Figure~\ref{fig:svedge}(b) where the saddle-node bifurcation is seen to
explain the dynamical  $I_{\rm left}$ for $f<0.29$.

However, it is the stability analysis of the $1000 \ldots 00$ state
that explains the expulsion of the vortex at the edge for $f>0.29$.
As shown in Figure~\ref{fig:svedge}(c), this configuration presents the
usual saddle-node ($I^{\star}_{\rm sv,left1}$) bifurcation. The pitchfork
($f^{\star}_{\rm sv,left1}$) bifurcation is barely visible
in the figure. There is also another saddle-node bifurcation at low $I$ and
high $f$ which is irrelevant for the depinning considered here.

The results of this section are summarized in
Figure~\ref{fig:svedge}(d), which indicates the region where {\it at least one}
single-vortex configuration in a ladder array {\it with edges} is dynamically
stable. Figure~\ref{fig:svedge}(d) is, in essence,
the union of Figure~\ref{fig:depinvortex}
with Figure~\ref{fig:svedge}(a)--(c) and shows how
the sv solutions either cease to exist through a saddle-node bifurcation
when $I$ is increased for most values of the frustration $f$, or
become unstable through a symmetry-breaking pitchfork bifurcation for
small values of $f$.

Within this picture, the dynamical behavior in
Figure~\ref{fig:depinvortex}, where the vortex propagates along the ladder
in the interval
$I^{\star}_{\rm sv,right}(f) < I < I^{\star}_{\rm sv,left}(f)$,
is the result of a succession of saddle-node bifurcations of
single-vortex states situated in contiguous cells until the vortex is
expelled at the left edge.
This is in contrast with the symmetry-breaking exit
of the vortex for $f < f^{\star}_{\rm sv}(I)$ where the flux is expelled
in the transversal direction.

This description is valid for $f<0.37$. However, a subtle variation is
observed beyond that point for the sv solutions. The comparison of
Figure~\ref{fig:svedge}(c) and Figure~\ref{fig:depanalyt} shows that
the depinning of the sv and nv states is similar for high $f$, i.e.
$I^{\star}_{\rm sv,left} \approx I^{\star}_{\rm nv}$, for $f>0.37$.
This is due to the fact that at high $f$ the vortex is {\it not} expelled
from the array before depinning. Instead, the sv configuration with
the vortex at the left edge ceases to exist through a saddle-node
bifurcation which, similarly to what happens to the nv state, is
localized on the {\it right} edge of the ladder. The depinning
transition of the single-vortex state at high f can be described in a
similar fashion as the depinning of the no-vortex configuration: by
the nucleation of vortices on the right edge of the ladder.
Thus, the single-vortex configuration plays no role in the
global depinning of the array: only the behavior of the nv and ff
states has to be considered.

\subsection{Bifurcations of the fully-frustrated state}

The depinning of the ladder array as $f \rightarrow 1/2$ is determined
by the stability of the fully-frustrated configuration. This can be
readily seen by inspection of Figure~\ref{fig:depinning} which shows that
at high $f$ the depinning of the no-vortex state occurs {\it before} the
global depinning of the array.

To clarify the importance of these transitions we follow the same scheme
as above. We first calculate the
zero-eigenvalue bifurcations of this state and show that the depinning
of the ladder at $f \simeq 1/2$ is indeed explained by a saddle-node
bifurcation of the ff state. Then we obtain analytical approximations
(``th'') to the exact bifurcations, using stability criteria for
the infinite-ladder ff configuration.

In principle, the characterization of the ff bifurcations is more
intricate than for the nv and sv configurations above since in 
the {\it finite} ladder several states of the ff type
could play a role in the depinning. First, 
there exist different states in ladders
with odd and even number of cells, as seen when the $\ldots 010101 \ldots$
alternating vortex pattern is fitted into a finite length.
Second, there are many ff-states very close energetically with 
different dynamic stability structure.
However, we will show that the landscape of relevant
solutions is indeed clear, and depinning can be assigned to instabilities
of {\it one} of those configurations.

Consider first an {\it even} ladder with ten plaquettes as an example.
Of the several states of the ff type, three can be thought as relevant:
(1a) $1010101010$, (1b) $0101010101$, and (1c) $1101010100$. We have analyzed
the stability of these three states and conclude that only configuration
(1a) is relevant for depinning of the even ladder. 
Both (1b) and (1c) cease to exist or become
unstable at lower critical currents. This is physically reasonable since
(1b) will tend to move one cell to the left under the action of
a driving current to produce (1a), while (1c) ceases to exist through
a low current saddle-node bifurcation caused by the expulsion of the vortex
at the left boundary.

The case of the {\it odd} ladder with $N$ plaquettes has one
further complication,
namely that we cannot have exactly $N/2$ vortices in the array. Thus, the
``pure'' ff state (as calculated for the infinite case) is not possible
under these topological constraints.
However, states similar to the ``pure'' ff are those which
contain $(N-1)/2$ and $(N+1)/2$ vortices. Take, for instance,
a ladder with eleven plaquettes. We have then two groups of states:
those with $6$ vortices,
(2a) $10101010101$, (2b) $11010101010$, and (2c) $01010101011$;
and those with $5$ vortices,
(3a) $10101010100$, (3b) $01010101010$, and (3c) $00101010101$.
The detailed analysis of these states shows that (3a) has the highest
critical transition and is thus responsible for the depinning of the
odd ladder.

\begin{figure}[t]
\psfig{file=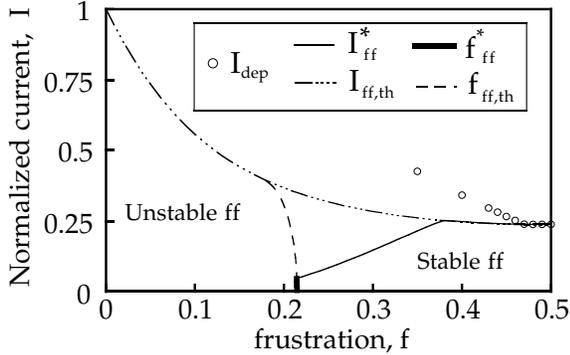,width=3.5in}
\caption{Dynamic stability of the fully-frustrated solution. The
saddle-node bifurcation at $I^{\star}_{\rm ff}$ is responsible for
the depinning of the array for $f \sim 0.5$.  The ff state
also undergoes a pitchfork bifurcation $f^{\star}_{\rm ff}$.
Both are partially explained through approximate formulas
$I_{\rm ff,th}$ and $f_{\rm ff,th}$ deduced from instabilities of
the ff solution for the {\it infinite} ladder.}
\label{fig:depff}
\end{figure}

There is a final important observation in our argument: 
the stability diagrams of the configurations which
cause the depinning in odd and even ladders --(3a) and (1a) respectively--
are {\it indistinguishable}. (This is also true for other odd-even
related configurations like (2b) and (1c).) The conclusion is then clear:
the depinning behavior of odd and even ladders at high $f$ is identical
as it can be explained by the bifurcation of configurations (1a) and (3a)
which are of the $101010 \ldots$ type.

The zero-eigenvalue bifurcations of these configurations are presented in
Figure~\ref{fig:depff}. The bifurcations are of two types (similar
to those obtained for the sv state):
saddle-node ($I^{\star}_{\rm ff}$) and pitchfork ($f^{\star}_{\rm ff}$).
Note also how the saddle-node bifurcation corresponds to two 
distinct dynamical instabilities: for $0.38< f <0.5$, the instability 
is spatially extended, while
for $0.22<f<0.38$, the instability is localized at the {\it right} edge.  
Only the former (spatially extended) saddle-node bifurcation has any 
relevance for the global depinning of the array---as reflected by 
the agreement between $I_{\rm dep}$ and $I^{\star}_{\rm ff}$ 
for $f > 0.45$.  (Of course, for $f < 0.45$, the
depinning current and $I^{\star}_{\rm ff}$ no longer coincide, because the
depinning is caused there by a saddle-node bifurcation of a different
state, the no-vortex solution, as shown earlier.)
The pitchfork bifurcation $f^{\star}_{\rm ff}$  is also spatially
extended and, as observed for the single vortex, it implies 
the breaking of the up-down symmetry in the ff
state. Thus, the flux is expelled in the transversal direction.
We clarify this in the following by calculating some analytical
approximations to these criteria.

\subsubsection{Analytical approximations to the ff bifurcations}

The rigorous analysis above indicates that two of the bifurcating
mechanisms imply spatially extended perturbations which are not
localized at the edges. Hence, we turn to instabilities of
the infinite-ladder fully-frustrated solution to obtain
analytical approximations.

One of the criteria was already established in Section~\ref{sec:analytical} 
where we showed that the no-edge ff solution ceases to exist at
a current $I_{\rm ff,th}(f)$ given by~(\ref{eq:currhalf}).
Indeed, we find excellent numerical agreement between the
analytical $I_{\rm ff,th}$ and the numerical
$I^{\star}_{\rm ff}$ for $0.38< f < 0.5$ (Figure~\ref{fig:depff}).
For instance, the predicted $I_{\rm ff,th}(f=1/2)= \sqrt 5 -2=0.236$ is
very close to the dynamically computed $I_{\rm dep}(f=1/2)=0.238$.

To explain the observed pitchfork bifurcation,
recall that the up-down symmetric manifold of solutions becomes unstable to
normal perturbations when the absolute value of the horizontal phases is
larger than $\pi/2$, as given in~(\ref{eq:horizontalcond}). Thus,
from~(\ref{eq:halfsolution})--(\ref{eq:Lhf}) and the critical condition
$|{\phi^{H}}^{\dagger}|=\pi/2$ we obtain the following implicit equation
for the instability boundary $f_{\rm ff,th}(I)$:
\begin{equation}
I=\tan (\pi f_{\rm ff,th}) \sqrt{\cos^{2} (\pi f_{\rm ff,th}) -
4 \sin^{4} (\pi f_{\rm ff,th})},
\label{eq:ffftrans}
\end{equation}
which is shown in Figure~\ref{fig:depff} and compared to the
exact $f^{\star}_{\rm ff}$ with excellent agreement.
As an example, it is easy to show analytically that the value of this
critical field when $I=0$ is given by
\begin{equation}
\label{eq:f_ff,th_zero}
f_{\rm ff,th}(I=0)=\frac{1}{\pi}
\arcsin \left[ \sqrt{\frac{17^{1/2}-1}{8}} \right ] = 0.2148.
\end{equation}

In summary, we can explain in part the stability diagram of the ff state
in the presence of edges with two bifurcations of the
infinitely extended (no-edge) ff solution: the saddle-node bifurcation
$I_{\rm ff,th}(f)$, and the subcritical pitchfork $f_{\rm ff,th}(f)$.

However, for the finite ladder, another
saddle-node bifurcation is reached before $I_{\rm ff,th}$
for $0.22< f < 0.38$ as seen in Figure~\ref{fig:depff}.
Numerical analysis shows that this bifurcation is local and it corresponds
to an instability in the {\it leftmost} cell. It can be approximated
heuristically by a criterion similar to (\ref{eq:heurcrit}) for the nv state,
i.e. for this interval of the frustration the ff state depins
approximately when the leftmost junction becomes unstable.

\section{Summary and discussion}
\label{sec:conclusion}

The first conclusion of our analysis is that for most values of $f$, the
global depinning current of the array $I_{\rm dep}(f)$
coincides with the current $I_{\rm nv}^{\star}$ where
the no-vortex state undergoes a saddle-node bifurcation
(Figure~\ref{fig:depinning}).
This bifurcation point can be well approximated by an analytical
$I_{\rm nv,th}$, given in~(\ref{eq:heuristic}), derived from
an instability criterion for the rightmost junction of the array.
For values of $f$ close to $1/2$, however,  the global depinning is caused
by a saddle-node bifurcation of the fully-frustrated solution at a current
$I_{\rm ff}^{\star}$. This bifurcation itself is well approximated 
by the global instability of
the no-edge ff state at $I_{\rm ff,th}$, as given analytically in
equation~({\ref{eq:currhalf}}).

We have also shown that the depinning of the
single vortex and its subsequent motion in the $\bf{- \hat x}$ direction
(Figure~\ref{fig:LAT}) is
the result of a cascade of saddle-node bifurcations of the single-vortex
solution such that, for most values of $f$, the fluxoids are expelled
from the ladder through its left edge.
More surprisingly, for smaller $f$ the sv and ff configurations
can also undergo another transition:
a symmetry-breaking subcritical pitchfork bifurcation in which
the up-down symmetry of the horizontal phases plays a crucial role.
In this case, the fluxoids
are expelled in the transversal ($\bf{- \hat y}$) direction through the
horizontal junctions. 

At a finer level of description, the approximations 
obtained in Section~\ref{sec:analytical}  for the
nv, ff and sv states all have a common feature: the
corrections due to the existence of edges, or of
topological vortices in the array, decay exponentially in space with a
characteristic length dependent on $I$ and $f$, 
as seen in~(\ref{eq:scnvr}) for instance.
Thus, the effect of the perturbations can usually be captured
by a local analysis.
This explains why, besides their independence from
the purely dynamical parameter $\beta_{c}$, the depinning observables are
largely independent of $N$, the length of the array.

\begin{figure}[t]
\psfig{file=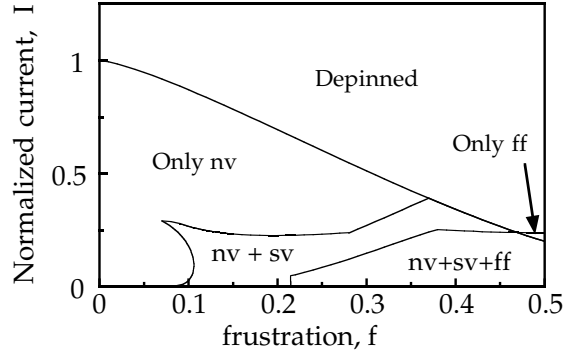,width=3.5in}
\caption{Dynamic stability diagram in the finite ladder (with edges)
of the superconducting states analyzed in this paper:
the no-vortex (nv), fully-frustrated (ff), and single-vortex (sv)
configurations. This figure combines 
Figures~\protect{\ref{fig:depanalyt}}, \protect{\ref{fig:svedge}}(d), and
\protect{\ref{fig:depff}}. 
Labels indicate the states that are {\it dynamically} stable
inside each region. In regions where two or more stable states coexist, 
each is attained from different initial conditions.}
\label{fig:phasediagedge}
\end{figure}

We have summarized our results in a zero-temperature stability diagram
(Figure~\ref{fig:phasediagedge}) where we present
the different critical currents
for the nv, sv, and ff superconducting solutions for the
{\it finite} ladder (with edges).
In short, the array ceases to be superconducting (depins globally)
at $I_{\rm dep}$ when either the no-vortex, single-vortex or fully-frustrated
solutions ceases to exist through saddle-node bifurcations.

The presence of vortices in the array does not change the observed depinning.
If $f<f_{\rm min}(I) \sim 0.1$, the single-vortex solution is always unstable.
For $f_{\rm min}<f<0.37$, a vortex in the array will depin at
$I_{\rm LAT} < I_{\rm dep}$ and will be expelled at the left edge
at $I_{\rm left} < I_{\rm dep}$. At that point
the no-vortex solution is recovered. This behavior is the same
for multivortex solutions with moderate $f$.
For $f>0.37$, the single-vortex is not expelled at the left edge 
before depinning but its instability is almost identical to that of 
the no-vortex configuration
since vortices enter from the right edge.

Note that no energy criteria have been
invoked above. The calculation of energy boundaries for the relevant
states remains open for further investigation.

\begin{figure}[t]
\psfig{file=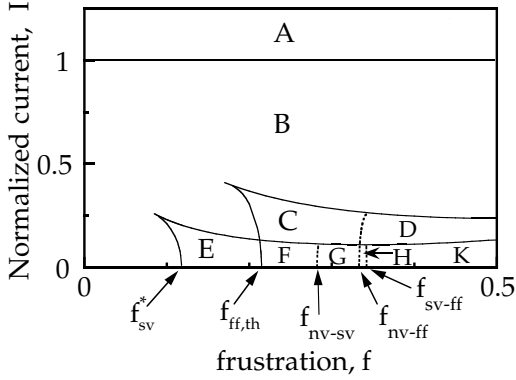,width=3.5in}
\caption{Phase diagram of the infinite ladder (no edges) for the
nv, sv and ff solutions. It can also be used to describe the ring ladder.
Solid lines are the dynamical critical currents derived in the
previous sections for solutions with no edges and for the vortex in the
center of the array. Dashed lines indicate the thermodynamic boundaries where
the energies of the approximate solutions are equal.
The physical meaning of the different regions can be summarized as follows:
A: Running solution (no superconducting solution exists);
B: Only nv exists, and it is stable;
C: $E_{\rm nv}<E_{\rm ff}$;
D: $E_{\rm ff}<E_{\rm nv}$;
E: $E_{\rm nv}<E_{\rm sv}$;
F: $E_{\rm nv}<E_{\rm sv} < E_{\rm ff}$;
G: $E_{\rm sv}<E_{\rm nv} < E_{\rm ff}$;
H: $E_{\rm sv}<E_{\rm ff} < E_{\rm nv}$;
K: $E_{\rm ff}<E_{\rm sv} < E_{\rm nv}$.
The dotted lines
$f_{\rm nv-sv}$, $f_{\rm nv-ff}$ and  $f_{\rm sv-ff}$ are approximate
thermodynamical criteria. In particular, $f_{\rm nv-sv}$ is analogue to the
$f_{c1}$ defined by Kardar~\protect{\cite{kardar2}} and
$f_{\rm nv-sv}(I=0) = 0.2823$ is in good agreement with his estimate.
Moreover, the zero-current energy boundary between
the nv and ff solutions~(\protect{\ref{eq:nvhfbound}}) can be shown
analytically to occur at $f_{\rm{nv-ff}}(I=0)=1/3$.
On the other hand, both $f^{\star}_{\rm sv}$ and
$f_{\rm ff,th}$ ensue from dynamical instability conditions.
For $I=0$, $f^{\star}_{\rm sv}$ can be approximated by
$f_{\rm {sv,th}}(I=0) = 0.1193$, as given
by~(\protect{\ref{eq:crithorizontal}}),
explaining the numerical observations
of Hwang {\it et al.}~\protect{\cite{stroudscreen}} for
$f_{c1 \perp}^{\ast}$. Another analytical
result~(\protect{\ref{eq:f_ff,th_zero}})
shows that $f_{\rm ff,th}(I=0)=0.2148$.}
\label{fig:phasediag}
\end{figure}

To highlight the effect of the edges and to include energetic considerations
we explore in more detail the phase diagram for the
infinite ladder (no edges), which we present in Figure~\ref{fig:phasediag}.
(This should be compared to the phase diagram of
the ladder with edges in Figure~\ref{fig:phasediagedge}.) 
When no edges are present, the dynamic bifurcation boundaries 
for the three states are
given by $I^{\dagger}_{\rm nv,th}$, $I_{\rm ff,th}$, $f_{\rm ff,th}$,
$I^{\star}_{\rm sv,center}$, and $f^{\star}_{\rm sv,center}$.
These dynamic criteria do not contradict previous
thermodynamic studies of the infinite
ladder~\cite{kardar1,kardar2} with $I=0$ where the no-vortex solution was
calculated to be energetically stable only for frustrations smaller
than a thermodynamic critical field $f_{c1}$, above which the
flux penetrates the ladder. We have extended these
calculations for the {\it driven} infinite ladder and the 
three superconducting solutions addressed in this article.
Defining the energy of a given configuration as
\[E= - \sum_{all} \cos \phi_{j},\]
we have calculated the energy boundaries for the approximate no-edge
no-vortex, fully-frustrated and single-vortex solutions:
$f_{\rm nv-sv}(I)$, $f_{\rm nv-ff}(I)$ and $f_{\rm sv-ff}(I)$. The results of
these {\it thermodynamic} calculations are also presented in
Figure~\ref{fig:phasediag}. Note, for instance, that our calculated
$f_{c1} = f_{\rm nv-sv}(I=0) \simeq 0.282$ agrees well with other
estimates~\cite{kardar2} $f_{c1}\simeq 2 \sqrt 2/\pi^{2}=0.287$.

Moreover, our approximate solutions produce some new
analytical results. For example, a closed expression for the energy
boundary between the no-vortex and
fully-frustrated solutions $f_{\rm {nv-ff}}(I)$ can be  obtained as:
\begin{eqnarray}
\sqrt{1-I^2} - I \sqrt{2/L - 1} + 2 \cos (\pi f_{\rm {nv-ff}}) \nonumber \\
- 2 \sin(\pi f_{\rm {nv-ff}}) \sqrt{1-2 I^{2}/L}=0
\label{eq:nvhfbound}
\end{eqnarray}
where $L$ is given by~(\ref{eq:Lhf}) with the negative sign.
For the special case $I=0$, and using the limit
$2I^{2}/L \rightarrow (1+4 \sin^{2} \pi f)^{-1}$,
it can be shown analytically that this boundary crosses the $I=0$ axis at
$f_{\rm{nv-ff}}(I=0)=1/3$.

We emphasize also that the regions in Figure~\ref{fig:phasediag}
present distinct dynamical and thermodynamical stabilities. For instance,
in some of them, the single-vortex solution is not the
ground state of the system, although it is {\it dynamically} stable
(metastable).

When comparing our results with those found in earlier work, one should
carefully note the direction of injection of the driving current $I$.
Previous analytical studies~\cite{kardar2,benedict}  have focused on the
$I=0$ case and considered the effect of a small {\it parallel} current in
the ${\bf \hat x}$ direction. In contrast, here
$I$ is injected in the {\it perpendicular} (${\bf{\hat y}}$) direction.
The depinning depends on the direction of current injection, a factor
to be taken into account when explaining recent numerical simulations of
ladder arrays~\cite{stroudscreen}. In those simulations,
marked differences between the depinning current of a circular ladder with
{\it perpendicular} injection, and of an open-ended ladder with {\it
parallel} injection were reported.

Although we have not studied the ring ladder in this paper,
it is closely connected to the infinite (no-edge) ladder.
The solutions for the infinite ladder constitute a
sub-manifold of the solutions for the ring ladder (i.e., the topological
constraints are more strict on the infinite ladder than in the ring ladder).
To understand this, note first that the governing
equations~(\ref{eq:eq1})--(\ref{eq:eq3}) are the same for both
ring and infinite ladders. On the other hand, following the same reasoning
given by the sequence of
equations~(\ref{eq:interiorequation1})--(\ref{eq:returnedges}), we conclude
that the topological constraints in the ring ladder imply only that
\begin{equation}
\label{eq:returnring}
I^{H}_{j}+I^{h}_{j}=C, \;\; \forall j,
\end{equation}
where $C$ is a constant. Therefore, from~(\ref{eq:returnedges}),
the infinite ladder is mathematically
equivalent to the particular case of the ring with $C=0$, in which
the concentric currents through the horizontal junctions in
the outer $I^{H}$ and inner $I^{h}$ circles are equal and opposite.

Consequently, the results for the infinite ladder summarized in
Figure~\ref{fig:phasediag} are also valid for the ring ladder if we
restrict to the submanifold with $C=0$. In this case, the phase diagram
has to be reinterpreted in terms of the new topological constraints.
First, the depinning of the no-vortex solution is unchanged:
$I^{\dagger}_{\rm nv,th}$ is still constant and equal to $1$.
However, the two bifurcations of
the single-vortex configurations have new dynamical meaning. If the sv
state goes unstable through the saddle-node bifurcation
$I^{\star}_{\rm sv} = I_{\rm LAT}$, flux cannot be expelled through the
horizontal junctions and the vortex depins and moves along the ladder
circularly. Thus, the ring ladder depins effectively at $I_{LAT}$.
However, if $f<f^{\star}_{\rm sv}$, the flux can
be expelled {\it transversally} from the ladder through the 
horizontal junctions
and the no-vortex state is recovered. Then, the depinning occurs at
$I^{\dagger}_{\rm nv,th}=1$. This is exactly the behavior reported from
numerical simulations by Hwang {\it et al.}~\cite{stroudscreen}. First,
their isotropic $I_{c \perp}$ coincides with our calculated
$I^{\star}_{\rm sv} = I_{\rm LAT}$. Second, they found a
critical field $f_{c1 \perp}^{\ast}
\simeq 0.12$ below which the depinning current is $I_{c \perp}=1$
with exclusion of field inside the array. This corresponds
to $f^{\star}_{\rm sv}$, the frustration
below which the single-vortex configuration becomes dynamically unstable
through a symmetry-breaking pitchfork bifurcation, which can be
approximated analytically by $f_{\rm sv,th}(I)$, as given
in~(\ref{eq:stabhorizontal}). (In the absence of driving, the analytical
prediction~(\ref{eq:crithorizontal}) gives $f_{\rm sv,th}(I=0)=0.1193$,
in perfect agreement with their numerical simulations~\cite{stroudscreen}.)

We note in passing that the same behavior should be expected for
the fully-frustrated state in the ring ladder.
The checkerboard pattern would begin to slide along the ring
at $I_{\rm ff,th}$ producing a finite voltage. However, if the
$f_{\rm ff,th}$ is crossed, the flux can be expelled transversally
and the nv state would appear. These predictions would have to be
checked numerically.

As a final remark, we also note that the second device considered by
Hwang {\it et al.}~\cite{stroudscreen}---an open-ended  ladder
with {\it parallel} current
injection---cannot be compared directly with our ladder with
{\it perpendicular} injection. However, the depinning
current $I_{c}(f)$ follows a similar trend to our $I_{\rm dep}(f)$.
In fact, the dependence of their $I_{c}(f)$ seems to be well explained
with formulae calculated by Benedict~\cite{benedict} for the same device
by invoking a similar criterion: the onset of soft modes~\cite{benedict2}.

\begin{acknowledgments}
We thank Shinya Watanabe, Enrique Tr\'{\i}as, and Herre van der Zant 
for consistently useful discussions, and Mehran Kardar for helpful 
guidance to the literature on ladder arrays.  Research supported
by a Spanish MEC-Fulbright predoctoral fellowship (MB), and by
NSF grants DMS-9500948 (SHS) and DMR-9402020 (TPO).
\end{acknowledgments}

\appendix
\section{Comparison of the single-vortex configuration in the ladder
with the one-dimensional kink}
\label{sec:appsg}

Although analogies between the ladder and strictly one-dimensional
parallel arrays can be drawn,
we show now how the single-vortex solution for the quasi-one-dimensional
ladder is mathematically different from
the kink-like vortex in 1-D parallel arrays.

Recall that the equations for the one-dimensional parallel array can be
reduced to a discrete driven sine-Gordon
equation~\cite{shilong,parallelcontinuum}
if only self-inductances are considered.
When $I=0$, the discrete single-vortex solution is well
approximated by the kink
solution of the undriven, time-independent, infinitely-extended,
one-dimensional continuum sine-Gordon equation~\cite{drazin}:
\begin{equation}
\label{eq:kink}
\phi^{\rm SG}_{i}=4 \arctan\{\exp[(i-i_{0})/\lambda_{\rm SG}]\}.
\end{equation}
Thus, the vortex  corresponds to a $2 \pi$-jump centered at $i_{0}$ with
a characteristic half-width $\lambda_{\rm SG}$.
(Incidentally, it has also been
shown~\cite{parallelcontinuum} that by introducing an effective
$\lambda_{\rm SG}^{\rm eff}$, this functional form is also valid
when mutual inductances are included).

For our no-inductance ladder array, it is also
possible to obtain an approximate sine-Gordon equation for the system.
Although there is no explicit inductance in the problem, the coupling
between the vertical junctions is provided by the
horizontal junctions via the fluxoid quantization.

The approximate sine-Gordon equation has been most simply
obtained~\cite{kardar2,strouddyn} by assuming
that the horizontal phases are small:
$\sin \phi^{H}_{j} \approx \phi^{H}_{j}$.
Then, the {\it zero-current} time-independent equations for the ladder
(\ref{eq:scequation1})--(\ref{eq:scequation2})
can be reduced to $\phi^{V}_{j+1}-2 \phi^{V}_{j} + \phi^{V}_{j-1} =
2 \sin \phi^{V}_{j}$. This gives, in the continuum limit,
the time-independent sine-Gordon equation with no forcing and
$\lambda_{\rm SG}=1/\sqrt{2}$,  where the cell size is taken as length unit.
However, a better linearization is suggested by the numerics if we take
the phase change in the vertical junctions
$(\phi^{V}_{j}-\phi^{V}_{j+1})/2 = \phi^{H}_{j} - \pi f \ll 1$ as
the small parameter. In other words, one should linearize about
$\phi^{H}_{j} = \pi f$, not $\phi^{H}_{j} = 0$.
In that case, we obtain the following more accurate
sine-Gordon equation: $\phi^{V}_{xx}-(2/\cos \pi f) \sin \phi^{V}=0$,
with $\lambda_{\rm SG}^2=\cos(\pi f)/2$.

\begin{figure}[t]
\psfig{file=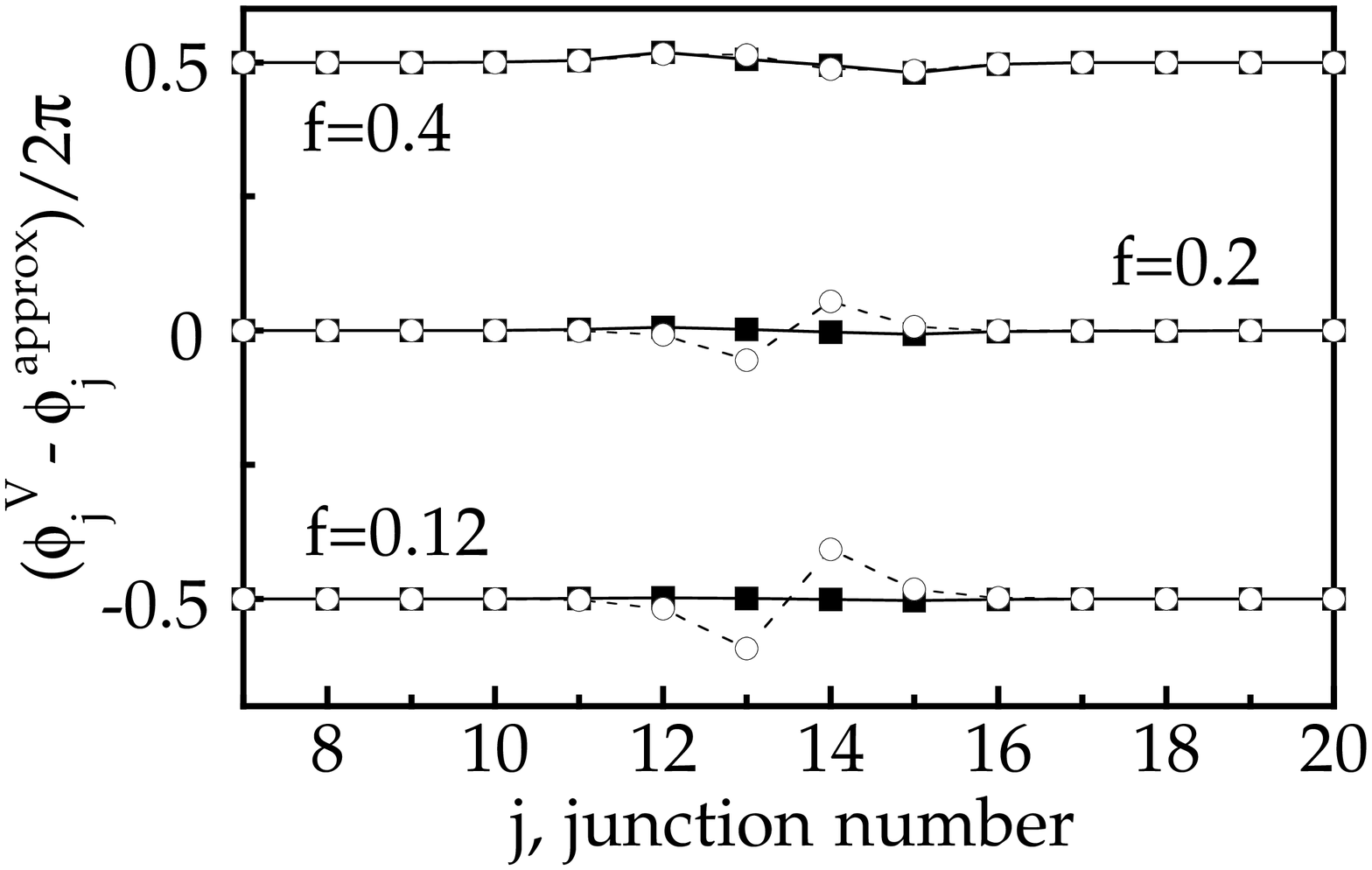,width=3.5in}
\caption{Comparison of the errors of our
approximation~(\protect{\ref{eq:onevortex}}) (black squares) and the
sine-Gordon kink~(\protect{\ref{eq:kink}}) (white circles)
for varying magnetic field far from the edges in a $25 \times 1$
array.  Graphs for different $f$ are offset by $\pi$ for
clarity.  The approximation~(\protect{\ref{eq:onevortex}}) has no
adjustable parameters.  In contrast,  to make the kink approximation as
accurate as possible, its characteristic length $\lambda_{\rm SG}$ was
chosen by a linear fit of $\ln \left[\tan (\phi^{V}_{j}/4) \right]$ vs.\
$j$, where the $\phi^{V}_{j}$ are the numerically computed phases of the
true single-vortex state.  Even with this {\it a posteriori} fit to the
numerical data, the kink is not as good an approximation
as~(\protect{\ref{eq:onevortex}}).}
\label{fig:sinegordon}
\end{figure}

We argue now that the numerically observed single-vortex configuration is
not as accurately approximated by the kink~(\ref{eq:kink}) as it is by
our expression~(\ref{eq:onevortex})--(\ref{eq:Bonevortex}).
To compare them,
we particularize~(\ref{eq:onevortex})--(\ref{eq:Bonevortex}) for $I=0$
for a long array such that $1 \ll a \ll N+1$. Then, the vertical phases
near the center of the vortex become
\begin{equation}
\label{eq:vortexlad}
\phi^{V}_{j}= \left \{
\begin{array}{ll}
P \exp\{(j-a)/\lambda\}, & j\leq a\\
2 \pi -P \exp\{(a+1-j)/\lambda\}, & j > a
\end{array}
\right .
\end{equation}
with $r$ and $\lambda$ given by~(\ref{eq:scnvr}) and $P$ by
\begin{equation}
\sin P=\sin\left(\pi f -\frac{r-1}{2r}P\right)+ \sin(\pi f +P).
\end{equation}
Note that in~(\ref{eq:vortexlad}) we have not reduced the phases to
the interval $[-\pi, \pi)$, to facilitate the comparison with~(\ref{eq:kink}).
Our solution resembles the sine-Gordon kink in that it describes a
$2 \pi$-jump with odd symmetry with respect to \mbox{ $x_{0}=a+1/2$}.
However, both the functional forms and the characteristic lengths
are different. Figure~\ref{fig:sinegordon} shows that the numerics
are better approximated over a wide range of $f$
by~(\ref{eq:vortexlad}) than by the sine-Gordon kink~(\ref{eq:kink}).

Given the relative inappropriateness of the sine-Gordon kink as a model of
the vortex in the no-inductance ladder array, we conclude that even for
the static case it is an oversimplification to
reduce the ladder to a one-dimensional parallel array where the horizontal
junctions are approximated by an effective inductance.  Other dynamic phenomena
observed in the ladder, e.g, the dynamical mechanism of retrapping from the
whirling mode,  reinforce this statement and will be discussed
elsewhere~\cite{usladder2}.

\section{No-vortex solution and depinning with self-inductance}
\label{sec:appendixa}

Following on the concepts and notation in
Sections~\ref{sec:analytical} and~\ref{sec:depinning}, we briefly consider
the ladder with self-inductance. This is a first approximation
to explore the effect of self-fields  on the no-vortex
solution and, consequently, on the global depinning current of the array.

When self-inductances are included, the time-independent
governing equations~(\ref{eq:scequation1})--(\ref{eq:scequation2}) become
\begin{eqnarray}
\label{eq:selfind1}
&I+\sin \phi^{H}_{j-1}=\sin \phi^{H}_{j}+\sin \phi^{V}_{j}\\
&\phi^{V}_{j}-\phi^{V}_{j+1}-2 \phi^{H}_{j}= 2 \pi (n_{j}-f)
- I_{j}^{m}/ \Lambda^2,
\label{eq:selfind2}
\end{eqnarray}
where $I_{j}^{m}$ is the mesh current in plaquette $j$ and
$\Lambda^2=L_{J}/L_{s}$ is 
the two-dimensional penetration depth and a measure of the discreteness of the
array. Due to the geometrical constraints of the ladder, it is readily seen
that $I_{j}^{m} = - I_{j}^{H}$ in this case.

For the no-vortex solution we still have
$\phi_{j}^{V} \simeq \phi_{j+1}^{V} = {\phi^{V}}^{\ast}$ and
$\phi_{j}^{H} \simeq \phi_{j+1}^{H}= {\phi^{H}}^{\ast}$ far from the edges.
From~(\ref{eq:selfind1}) we then get
\begin{equation}
\label{eq:selfindvertical}
{\phi^{V}}^{\ast}=\arcsin I,
\end{equation}
as for the case with no inductance ($\Lambda = \infty$).
However, the horizontal phase is different and
has to be calculated from the following nonlinear equation:
\begin{equation}
\label{eq:selfindhorizontal}
{\phi^{H}}^{\ast}+\frac{1}{2\Lambda^2}\;{\sin {\phi^{H}}^{\ast}} = \pi f.
\end{equation}
This is a particular case of Kepler's equation, studied in
celestial mechanics~\cite{celestial}, which can be solved through the
method of successive approximations. In our case,
a good approximation to ${\phi^{H}}^{\ast}$ over the whole range of $f$
is given by the first iteration of that method as:
\[{\phi^{H}}^{\ast} \simeq \pi f - \frac{1}{2 \Lambda^2} \;
\sin \left(\frac{2 \Lambda^2 }{1+ 2 \Lambda^2} \; \pi f\right).\]

The procedure to account for the open boundaries
is identical to that presented in Section~\ref{sec:analytical}.
The results are also similar: the corrections decay exponentially from
the edges as in~(\ref{eq:correctionhor})--(\ref{eq:scnvr}).
However, the characteristic length $\lambda_{s}=1/\ln r_{s}$ is now
given by $r_{s}= \alpha_{s} + \sqrt{\alpha_{s}^{2}-1}$ where
\begin{equation}
\label{eq:selfindalpha}
\alpha_{s}=1+\frac{\sqrt{1-I^{2}}}{\cos{\phi^{H}}^{\ast}} \;
\left ( 1 + \frac{1}{2 \Lambda^2} \cos{\phi^{H}}^{\ast}\right ).
\end{equation}

\begin{figure}[t]
\psfig{file=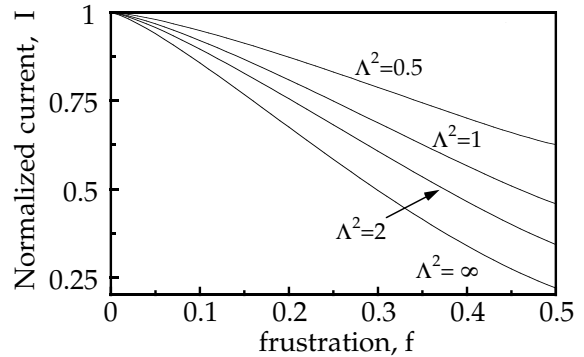,width=3.5in}
\caption{Effect of the self-inductance on the depinning current.
We show how the heuristic depinning approximation
$I_{\rm nv,th}(f)$,
obtained from~(\protect{\ref{eq:selfinddepin}}),
varies with the self-inductance $\Lambda^2 = L_{J}/L_{s}$. 
Note that the rest of the
paper deals with the limiting case $\Lambda =\infty$ where all inductances
are neglected.}
\label{fig:selfinductance}
\end{figure}

To assess how the inductance affects the depinning
current, we use~(\ref{eq:selfindalpha}) and the ${\phi^H}^{\ast}$
obtained from (\ref{eq:selfindhorizontal})
with a heuristic criterion similar to~(\ref{eq:heuristic}),
\begin{equation}
\label{eq:selfinddepin}
\arcsin (1-I_{\rm nv,th}) = {\phi^{H}}^{\ast} -
\frac{r_{s}-1}{2 r_{s}} \arccos I_{\rm nv,th},
\end{equation}
to calculate $I_{\rm nv,th}(f,\Lambda^2)$.
 The results in Figure~\ref{fig:selfinductance} show that, for fixed
$\Lambda$, $I_{\rm nv,th}(f)$ still decreases
monotonically with $f$. However, for a given $f$, the value of the 
depinning current increases as $\Lambda$ diminishes.
This is expected on physical grounds since an increase in the
inductance $L_{s}$ implies larger self-fields which oppose
the external applied magnetic field, thus decreasing the effective
magnetic flux through the array.


\newpage

\Large{Erratum: Superconducting states and depinning transitions of 
Josephson ladders [Phys. Rev. B 57, 1181 (1998)]}

\large{Mauricio Barahona, Steven H. Strogatz, and Terry P. Orlando}

\vspace{.3in}
The first formula of Section VI should include a contribution from
the driving current. In our choice of gauge,
\[E= - \sum_{\rm{all}} \cos \phi_{j} - I \sum_{j=1}^{N+1} \phi^{V}_{j}.\]
Accordingly, Eq.~(64) becomes
\[
\sqrt{1-I^2} + 2 \cos (\pi f_{\rm {NV-FF}})+I \arcsin(I)
- 2 \sin(\pi f_{\rm {NV-FF}}) (1-2 I^{2}/L)^{-1/2}
- I \arcsin(\sqrt{L/2})=0, \]
and $f_{\rm{NV-FF}}(I=0)=1/3$ remains unaltered.
Figure~15 is very slightly modified as shown. The conclusions do
not change.

\begin{figure}[t]
\psfig{file=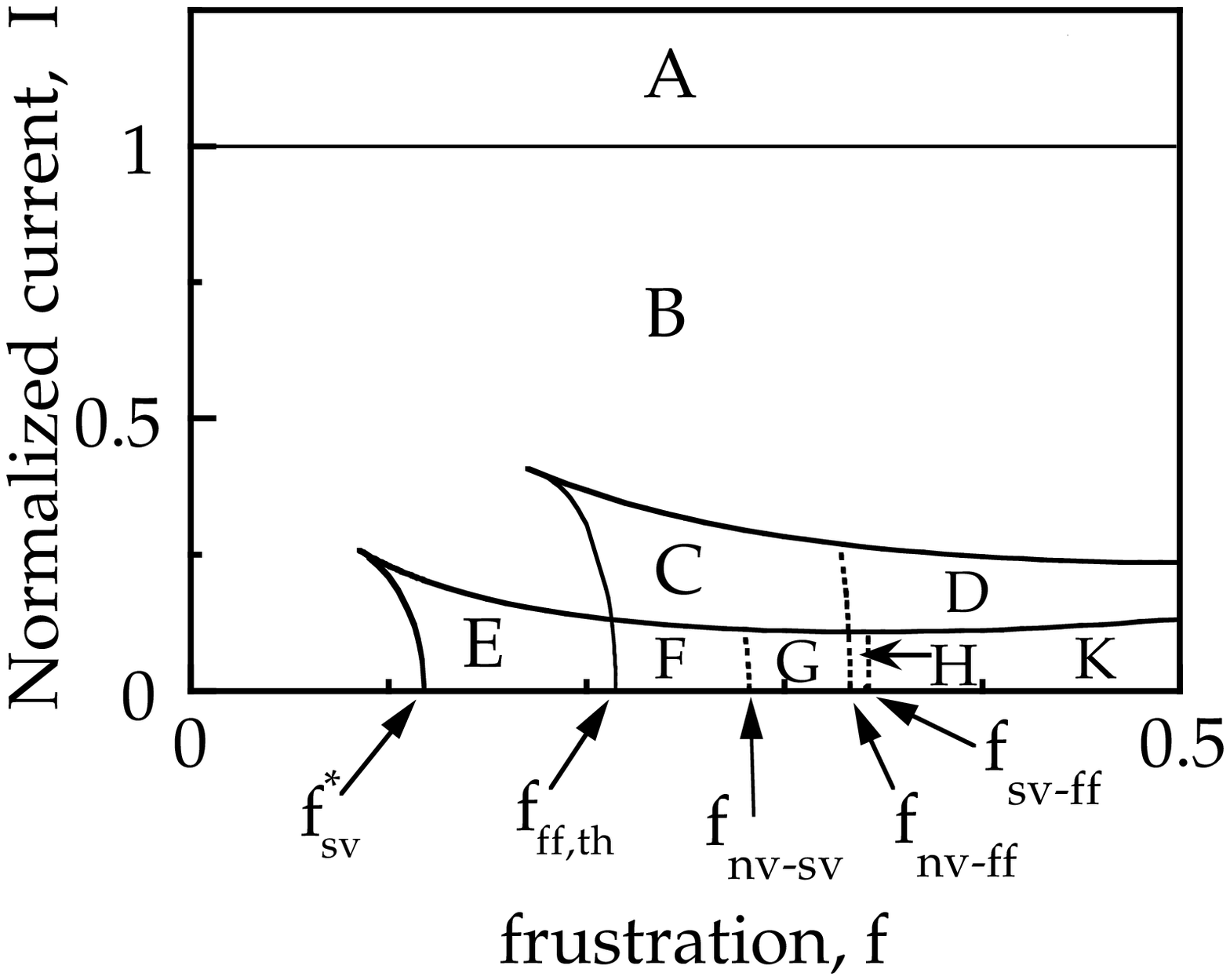,width=3.5in}
\vspace{-1in}
\caption{ (Corrected Fig. 15) Phase diagram of the NV, SV and FF 
solutions in the 
infinite (no edges) ladder---also relevant for the ring ladder.
Note that region H shrinks as the number of junctions is 
increased (i.e.\ $f_{\rm sv-ff}$ tends asymptotically to $f_{\rm nv-ff}$ as 
$N \to \infty$). Thus, the region H in the figure is only indicative.}
\label{fig:corrfig15}
\end{figure}


\begin{references}

\bibitem{scbooks} K.K. Likharev, {\it Dynamics of Josephson Junctions
and Circuits}, (Gordon and Breach, New York, 1986);
T. van Duzer and C.W. Turner, {\it Principles of
Superconductive Devices and Circuits} (Elsevier, New York, 1981);
M. Tinkham, {\it Introduction to Superconductivity}
(2$^{\rm nd}$ ed.) (McGraw-Hill, New York, 1996);
T.P. Orlando and K.A. Delin, {\it Foundations of Applied
Superconductivity}, (Addison-Wesley, Reading, MA, 1991).

\bibitem{technology} P.A.A. Booi and S.P. Benz, Appl. Phys. Lett. {\bf 64},
2163 (1994); S.P. Benz and C.A. Hamilton, Appl. Phys. Lett. {\bf 68},
3171 (1996).

\bibitem{kleiner} R. Kleiner {\it et al.}, Phys. Rev. Lett. {\bf 68}, 2394
(1992); R. Kleiner {\it et al.}, Phys. Rev. B {\bf 50}, 3942 (1994).

\bibitem{mpa1} R.A. Hyman {\it et al.}, Phys. Rev. B {\bf 51}, 15304 (1995).

\bibitem{series} P. Hadley, M.R. Beasley, and K. Wiesenfeld, Phys. Rev B
{\bf 38}, 8712 (1988);
A.A. Chernikov and G. Schmidt, Phys. Rev. E {\bf 52}, 3415 (1995);
K. Wiesenfeld, P. Colet, and S.H. Strogatz, Phys. Rev. Lett. {\bf 76}, 404
(1996);
S. Watanabe and J. Swift, {\it J. Nonlinear Sci.}, to be published.

\bibitem{shiseries} S. Watanabe and S.H. Strogatz, Physica (Amsterdam)
{\bf 74D}, 197 (1994).

\bibitem{shilong} H.S.J. van der Zant {\it et al.},
Phys. Rev. Lett. {\bf 74}, 174 (1995);
S. Watanabe {\it et al.}, Phys. Rev. Lett. {\bf 74}, 379 (1995);
S. Watanabe {\it et al.}, Physica (Amsterdam) {\bf 97D}, 429 (1996).

\bibitem{summary} A good sample of current research is presented in
{\it Macroscopic Quantum Phenomena and Coherence in Superconducting
Networks}, ed. C. Giovanella and M. Tinkham (World Scientific, Singapore,
1995). See also Physica (Amsterdam) {\bf 222B}(4), 253-406 (1996).

\bibitem{kardar0} M. Kardar and D. Erta\c{s} in {\it Scale Invariance,
Interfaces, and Non-Equilibrium Dynamics}, NATO ASI Series (Plenum,
New York, 1995).

\bibitem{steve} S.H. Strogatz, {\it Nonlinear Dynamics and Chaos: With
Applications to Physics, Biology, Chemistry, and Engineering}
(Addison-Wesley, Reading, MA, 1994), specially pp. 241-253.

\bibitem{experIV} J.S. Chung, K.H. Lee, and D. Stroud, Phys. Rev. B {\bf 40},
6570 (1989); F. Falo, A.R. Bishop, and P.S. Lomdahl, Phys. Rev. B {\bf 41},
10983 (1990); D. Reinel {\it et al.}, Phys. Rev. B {\bf 49}, 9118 (1994);
J.R. Phillips, H.S.J. van der Zant, and T.P. Orlando,
Phys. Rev. B {\bf 50}, 9380 (1994); D. Dom\'{\i}nguez and J.V. Jos\'e,
Mod. Phys. B {\bf 8}, 3749 (1994).

\bibitem{analysis2d} 
I.F. Marino and T.C. Halsey, Phys. Rev. B {\bf 50}, 6289 (1994); 
A.S. Landsberg, Y. Braiman, and K. Wiesenfeld, Phys. Rev. B
{\bf 52}, 15458 (1995); 
G. Filatrella and K. Wiesenfeld, J. Appl. Phys. {\bf 78}, 1878 (1995); 
I.F. Marino, Phys. Rev. B {\bf 52}, 6775 (1995);
I.F. Marino,  Phys. Rev. B  {\bf 55}, 551 (1997);
M. Barahona {\it et al.}, Phys. Rev. B  {\bf 55}, R11989 (1997).


\bibitem{kardar1} M. Kardar, Phys. Rev. B {\bf 30}, 6368 (1984).

\bibitem{kardar2} M. Kardar, Phys. Rev. B {\bf 33},  3125 (1986).

\bibitem{benedict} K.A. Benedict, J. Phys.: Condens. Matter {\bf 1}, 4895
(1989).

\bibitem{falo2} J.J. Mazo, F. Falo, and L.M. Flor\'{\i}a, Phys. Rev. B
{\bf 52}, 10433 (1995).

\bibitem{denniston} C. Denniston and C. Tang, Phys. Rev. Lett. {\bf 75},
3930 (1996).

\bibitem{stroudscreen} I-J. Hwang, S. Ryu, and D. Stroud, Phys. Rev. B
{\bf 53}, R506 (1996).

\bibitem{strouddyn} S. Ryu, W. Yu, and D. Stroud, Phys. Rev. E {\bf 53}, 2190
(1996); W. Yu, E.B. Harris, S.E. Hebboul, J.C. Garland, and D. Stroud,
Phys. Rev. B {\bf 45}, 12624 (1992).

\bibitem{mythesis} M. Barahona, Ph.D. Thesis (MIT, June 1996).

\bibitem{usladder2} M. Barahona, S.H. Strogatz, and T.P. Orlando,
to be published.

\bibitem{transverse}
J.F. Heagy, T.L. Carroll, and L.M. Pecora, Phys. Rev. E {\bf 50}, 1874 (1994);
E. Ott and J.C. Sommerer, Phys. Lett. A {\bf 188}, 39 (1994);
P. Ashwin, J. Buescu, and I. Stewart, Phys. Lett. A {\bf 193}, 126 (1994);
Y.-C. Lai and C. Grebogi, Phys. Rev. E {\bf 52}, R3313 (1995);
P. Ashwin, J. Buescu, and I. Stewart, Nonlinearity {\bf 9}, 703 (1996);
L. Kocarev and U. Parlitz, Phys. Rev. Lett. {\bf 76}, 1816 (1996);
R. Brown and N.F. Rulkov, Chaos (submitted).

\bibitem{parallelcontinuum} 
H.S.J. van der Zant, E.H. Visscher, D.R. Curd,
T.P. Orlando, and K.A. Delin, IEEE Trans. Appl. Supercond. {\bf 3},
2658 (1993); 
R.D. Bock, J.R. Phillips, H.S.J. van der Zant, and T.P. Orlando,
Phys. Rev. B {\bf 49}, 10009 (1994).

\bibitem{benednote} The other solution
for~(\ref{eq:halfsolution})--(\ref{eq:halfsolution2})
at $I=0$, $\{ {\phi^{V}_{j}}^{\dagger}=\pi/2 \; [1+(-1)^{j}], \;
{\phi^{H}_{j}}^{\dagger}=0 \}$ is unstable and not a ground state
of the system.

\bibitem{LAT} C.J. Lobb, D.W. Abraham, and M. Tinkham, Phys. Rev. B {\bf 27},
150 (1983).

\bibitem{terryvortex} H.S.J. van der Zant, T. P. Orlando, and J.E. Mooij,
Phys. Rev. B {\bf 43}, 10218 (1991).

\bibitem{newpaper} M. Barahona and S.H. Strogatz, to be published.

\bibitem{calculus} W. Kaplan, {\it Advanced Calculus} (Addison-Wesley, 
Reading, MA, 1984), pp.\ 117-120.

\bibitem{benedict2} The soft-mode criterion is mathematically
equivalent to a zero-eigenvalue bifurcation but does not 
identify Hopf bifurcations.
Since none of the transitions described in
this paper is Hopf, the soft-mode condition is equivalent to our dynamical
instability criterion given in~(\ref{eq:augmented2}).

\bibitem{drazin} P.G. Drazin, {\it Solitons: An Introduction}
(Cambridge University Press, New York, 1989).

\bibitem{celestial} P.M. Fitzpatrick, {\it Principles of Celestial Mechanics}
(Academic Press, New York, 1970), pp.\ 68-77.


\end{references}
\end{document}